\documentclass[a4paper,hidelinks, 12pt]{article}
\usepackage[utf8]{inputenc}
\usepackage{a4wide}
\usepackage{graphics}
\usepackage{amsmath}
\usepackage{amsfonts}
\usepackage{amssymb}

\usepackage{subfigure}
\usepackage{cite}
\usepackage{xcolor}
\usepackage{soul}
\usepackage{cancel}
\usepackage[toc,page]{appendix}
\numberwithin{equation}{section}
\usepackage{mathtools}
\usepackage{hyperref}
\usepackage{enumitem}
\usepackage[normalem]{ulem}

\allowdisplaybreaks

\newcommand{\hc}{{\rm h.c.}}

\def\gsim{\mathrel{\raise.3ex\hbox{$>$\kern-.75em\lower1ex\hbox{$\sim$}}}}

\newcommand{\diag}{{\rm diag}}

\begin{document}
\begin{titlepage}
\begin{center}

{\large \bf {Dark matter in three-Higgs-doublet models with \boldmath$S_3$ symmetry}}

\vskip 1cm

W. Khater, $^{a,}$\footnote{E-mail: Wkhater@birzeit.edu}
A. Kun\v cinas, $^{b,}$\footnote{E-mail: Anton.Kuncinas@tecnico.ulisboa.pt}
O. M. Ogreid,$^{c,}$\footnote{E-mail: omo@hvl.no}
P. Osland$^{d,}$\footnote{E-mail: Per.Osland@uib.no} and 
M. N. Rebelo$^{b,}$\footnote{E-mail: rebelo@tecnico.ulisboa.pt}

\vspace{1.0cm}

$^{a}$Birzeit University, Department of Physics,\\ P.O. Box 14, Birzeit, West Bank, Palestine,\\
$^{b}$Centro de F\'isica Te\'orica de Part\'iculas, CFTP, Departamento de F\'\i sica,\\ Instituto Superior T\'ecnico, Universidade de Lisboa,\\
Avenida Rovisco Pais nr. 1, 1049-001 Lisboa, Portugal,\\
$^{c}$Western Norway University of Applied Sciences,\\ Postboks 7030, N-5020 Bergen, 
Norway, \\
$^{d}$Department of Physics and Technology, University of Bergen, \\
Postboks 7803, N-5020  Bergen, Norway
\end{center}

\vskip 3cm

\begin{abstract}
Models with two or more scalar doublets with discrete or global symmetries can have vacua with vanishing vacuum expectation values in the bases where symmetries are imposed. If a suitable symmetry stabilises such vacua, these models may lead to interesting dark matter candidates, provided that the symmetry prevents couplings among the dark matter candidates and the fermions.  We analyse three-Higgs-doublet models with an underlying $S_3$ symmetry. These models have many distinct vacua with one or two vanishing vacuum expectation values which can be stabilised by a remnant of the $S_3$ symmetry which survived spontaneous symmetry breaking.
We discuss all possible vacua in the context of $S_3$-symmetric three-Higgs-doublet models, allowing also for softly broken $S_3$, and explore one of the vacuum configurations in detail.
In the case we explore, only one of the three Higgs doublets is inert. The other two are active, and therefore the active sector, in many aspects, behaves like a two-Higgs-doublet model. The way the fermions couple to the scalar sector is constrained by the $S_3$ symmetry and is such that the flavour structure of the model is solely governed by the $V_\text{CKM}$ matrix which, in our framework, is not constrained by the $S_3$ symmetry. This is a key requirement for models with minimal flavour violation. In our model there is no CP violation in the scalar sector.
We study this model in detail giving the masses and couplings and identifying the range of parameters that are compatible with theoretical and experimental constraints, both from accelerator physics and from astrophysics.
\end{abstract}

\end{titlepage}

\section{Introduction}
Cosmological observations, based on the standard cosmological model, 
$\Lambda_\mathrm{CDM}$, where CDM stands for cold dark matter, indicate that around a quarter of the total mass-energy density of the Universe is made up of Dark Matter (DM) \cite{Planck:2018vyg}.

In this paper a DM model based on the $S_3$-symmetric three-Higgs doublet model (3HDM) potential is studied in detail. In our framework the stability of the DM sector results from a  $\mathbb{Z}_2$ symmetry which survives the spontaneous symmetry breakdown of the initial $S_3$ symmetry. One of the doublets provides the DM sector, while the other two are active. Therefore the active sector behaves in many ways like a two-Higgs doublet model (2HDM)~\cite{Gunion:1989we,Branco:2011iw}.

The paper is organised as follows. In section~\ref{sect:Intro_2} a brief overview of the simplest implementations of scalar DM is presented, together with a list of references.  First, we mention the Inert Doublet Model (IDM)~\cite{Deshpande:1977rw,Barbieri:2006dq}, a 2HDM  
which constitutes  one of the first attempts at accounting for DM through the Higgs portal. Next we comment on implementations within 3HDMs. In section~\ref{sect:S3-potential} we introduce the $S_3$-symmetric potential, and discuss the different vacua~\cite{Emmanuel-Costa:2016vej} starting from a real scalar potential. It has 
been shown that the $S_3$-symmetric potential allows for spontaneous CP violation \cite{Emmanuel-Costa:2016vej,Ogreid:2017alh}.
In this section we give the motivation to study a particular vacuum, denoted R-II-1a, on which the rest of our paper is based. With this choice of vacuum there is no CP violation in the scalar sector. Next, in section~\ref{sect:R-II-1a}, we give the spectrum of masses of both the inert and the active sector of this model, specifying the rotations leading to the physical scalars, together with the scalar gauge couplings and the Yukawa couplings. The scalar-scalar couplings are listed in Appendix A. 

The theoretical and experimental constraints are imposed on the model in section~\ref{sect:exp-constraints}. This is done step-by-step by imposing in succession a series of cuts and by performing the corresponding numerical analysis. We start with Cut 1: perturbativity, stability, unitarity checks, LEP constraints together with the value of the recently measured Higgs boson mass \cite{ATLAS:2012yve,CMS:2012qbp}
by plotting the allowed regions for masses and several of the parameters of the model. Next, we apply Cut 2: SM-like (Standard Model) gauge and Yukawa sector, electroweak precision observables and $B$ physics; which further restrict the model as illustrated in that section. Finally, Cut 3 takes into account the limits imposed on the SM-like Higgs boson, $h$, by the decays $h \to \{\mathrm{invisible},~\gamma \gamma \}$, the DM relic density, and direct searches. The micrOMEGAs code \cite{Belanger:2008sj,Belanger:2013oya,Barducci:2016pcb} is used at this stage to evaluate the cold dark matter relic density along with the decay widths and other astrophysical observables. The full impact of Cut 3 is discussed in section~\ref{sect:Cut_3}, where we also present a set of benchmarks for this model. Two different scenarios emerge right from the beginning of the numerical analysis presented in section~\ref{sect:exp-constraints}, based on the ordering of the masses of the two neutral inert scalars, $\eta$ and $\chi$, which have opposite CP parities. The origin of the masses of these two fields is also different in terms of parameters of the potential. Cut 3 removes the possibility of having the $\eta$ scalar lighter than $\chi$. As a result, in this model only  $\chi$ can play the r\^ ole of DM. We also conclude that there are no good DM candidates for high mass values as is explained in our discussion. In section~\ref{sect:conclude} we present our conclusions. There are several Appendices where features of the model and details of our analysis are given.

\section{Scalar dark matter}
\label{sect:Intro_2}
\setcounter{equation}{0}

One of the simplest extensions of the SM which could accommodate DM is obtained by adding a scalar singlet. With an explicit $\mathbb{Z}_2$ symmetry, to prevent specific decay channels, this extension could yield a viable DM candidate \cite{Silveira:1985rk,McDonald:1993ex,Burgess:2000yq,Patt:2006fw,Barger:2007im,Andreas:2008xy}. Direct detection constraints were studied \cite{Mambrini:2011ik,Low:2011kp,Djouadi:2011aa,He:2011gc} based on the LHC data. The parameter space of these models was further constrained after the Higgs boson discovery \cite{Cline:2013gha,Feng:2014vea,Beniwal:2015sdl,Cuoco:2016jqt,He:2016mls,Casas:2017jjg,Athron:2017kgt}. While being the subject of specific exclusion criteria, in general, two DM regions were identified: a low-mass region 55--63~GeV and a high-mass region above around 100-500~GeV, depending on the implementation. In alternative, the scalar singlet DM can be stabilised by a different symmetry, such as $\mathbb{Z}_3$, which was considered in Refs.~\cite{Belanger:2012zr,Ko:2014nha}.

A DM candidate can also be introduced through non-trivial SU(2)$_L$ scalar $n$-tuples. In Ref.~\cite{Cirelli:2005uq} a class of so-called minimal dark matter (MDM) multiplets was proposed. The key aspect of these models is to extend the SM by additional scalar multiplets with minimal quantum numbers (spin, isospin, hypercharge) to accommodate the DM candidate. The MDM models were further on studied in Refs.~\cite{Cirelli:2007xd,Cirelli:2008id,Cirelli:2009uv,Hambye:2009pw,Cai:2012kt,Earl:2013jsa,Garcia-Cely:2015dda,DelNobile:2015bqo}.

\subsection{The Inert Doublet Model}
One of the most popular models, capable of accommodating DM, is the IDM \cite{Deshpande:1977rw,Barbieri:2006dq}. The IDM accommodates DM in the form of a neutral scalar: the lightest neutral member of an inert SU(2) Higgs doublet added to the SM. In addition to providing an economical accommodation of DM, the IDM also offers a mechanism for generating neutrino masses \cite{Ma:2006km,Kubo:2006yx,Andreas:2009hj}. The IDM was studied extensively some ten years ago \cite{Majumdar:2006nt,LopezHonorez:2006gr,Gustafsson:2007pc,Hambye:2007vf,Cao:2007rm,Lundstrom:2008ai,Agrawal:2008xz,Nezri:2009jd,Hambye:2009pw,Dolle:2009fn,Arina:2009um,Dolle:2009ft,Honorez:2010re,Miao:2010rg,LopezHonorez:2010tb,Sokolowska:2011sb}. Assuming the Higgs mass to be around 120~GeV, two DM mass regions were identified: a low and intermediate-mass region, from about 5~GeV to around 100~GeV (allowing for a heavy Higgs up to 500~GeV, this DM mass region would extend up to 160~GeV), and a high-mass region, beyond about 535~GeV. Above some 80~GeV, the annihilation in the early Universe to two gauge bosons ($W^+W^-$ or $ZZ$) becomes very fast and the relic DM density would be too low. In addition, annihilation into two SM-like Higgs bosons is possible and is controlled by the effective $\lambda_L$ coupling (parameterising the Higgs-DM-DM coupling), for small $\lambda_L$ values the overall effect is negligible. Eventually, for sufficiently heavy DM particles, above approximately 500--535~GeV, the annihilation rate drops and as a result the DM density is again compatible with the data.  

When the mass splitting between the DM candidate and the other scalars in the inert sector is sufficiently small, coannihilation effects get stronger. As a result, if the Higgs-DM-DM coupling, $\lambda_L$, has a suitable value, a relic density in agreement with observation can be achieved. For a fixed mass splitting with an increasing DM mass the absolute value of the $\lambda_L$ coupling should increase to satisfy the experimental relic density value. On the other hand, for a fixed $\lambda_L$ value, if the mass splittings are increased, the relic density decreases. The relic density depends also on the sign of $\lambda_L$.  Interplay of these  parameters may result in an acceptable relic density for the DM candidate of several TeV.

\begin{figure}[htb]
\begin{center}
\includegraphics[scale=0.55]{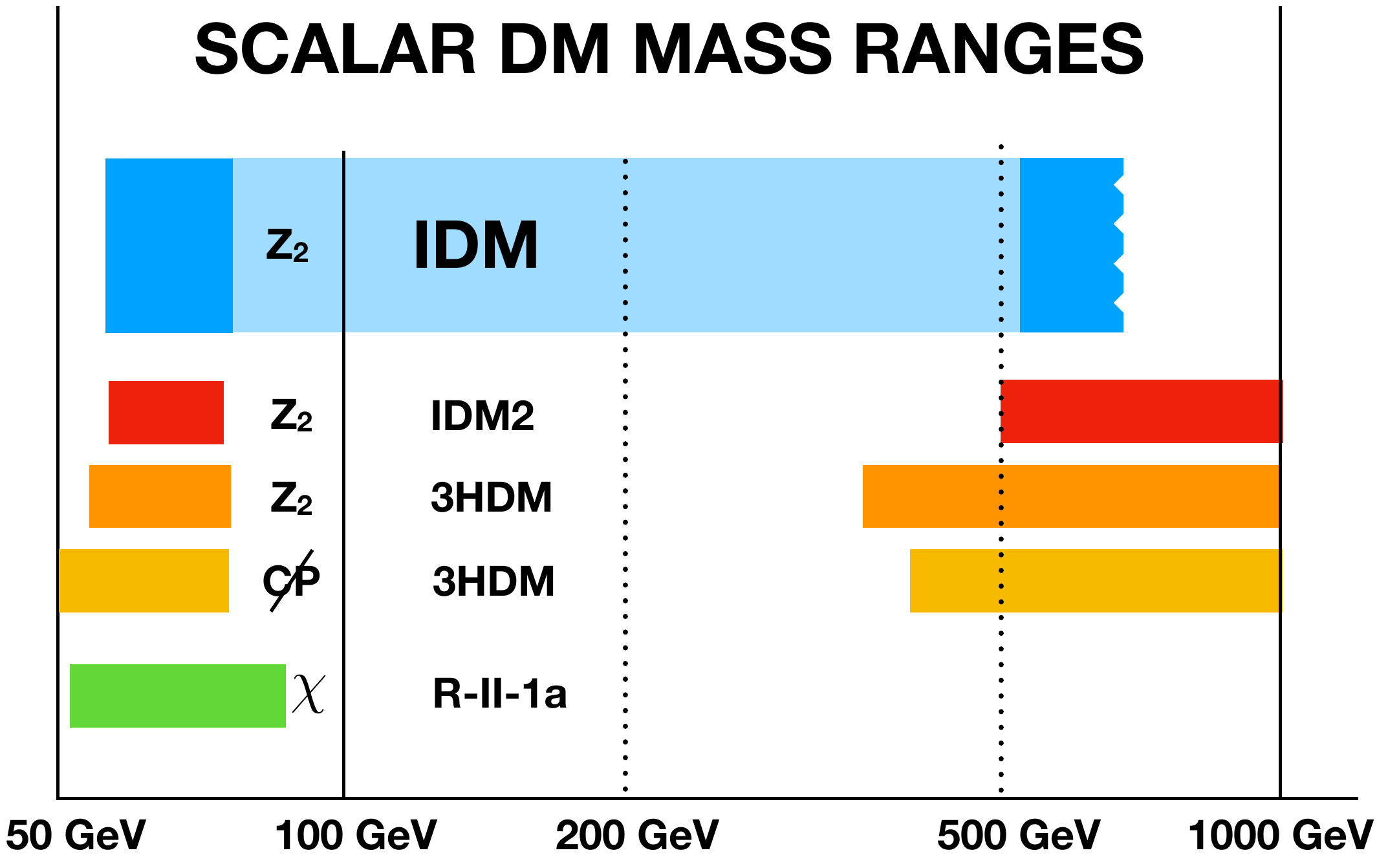}
\end{center}
\vspace*{-4mm}
\caption{ Sketch of allowed DM mass ranges up to 1~TeV in various models. Blue: IDM according to Refs.~\cite{Belyaev:2016lok,Kalinowski:2018ylg}, the pale region indicates a non-saturated relic density. Red: IDM2 \cite{Merchand:2019bod}. Ochre: 3HDM without \cite{Keus:2014jha,Keus:2015xya,Cordero:2017owj} and with CP violation \cite{Cordero-Cid:2016krd}. Green: the R-II-1a $\chi$ model presented in this paper.}
\label{Fig:mass-ranges}
\end{figure}

With the Higgs boson discovery~\cite{ATLAS:2012yve,CMS:2012qbp}, it was pointed out that the decay width of the Higgs boson to two photons and invisible particles could further constrain the IDM \cite{Swiezewska:2012eh,Goudelis:2013uca,Krawczyk:2013jta,Arhrib:2013ela}. It was found \cite{Ilnicka:2015jba,Diaz:2015pyv} that with both astroparticle and collider constraints taken into account the DM mass region below 45--50~GeV is ruled out, and thus a minimal mass threshold is set. Later, the whole low- and intermediate-mass region was shown to be consistent with observations for the DM candidate with masses 55--74~GeV \cite{Belyaev:2016lok}. Although the high-mass region, above 500~GeV, requires severe fine-tuning it is still compatible with data. The region with the DM masses 74--500~GeV is still possible: the exclusion of a given mass region in the parameter space depends on whether one requires the IDM to provide a single component DM candidate to generate all relic density or just partially contribute to the DM relic density. More recently, Ref.~\cite{Kalinowski:2018ylg} confirmed earlier mass-range observations of the IDM and provided a set of benchmarks for $e^+e^-$ studies, for IDM detection at potential $e^+e^-$ colliders (ILC, CLIC) see Refs.~\cite{Aoki:2013lhm,Hashemi:2015swh,Kalinowski:2018kdn}. The aforementioned DM mass ranges are sketched in figure~\ref{Fig:mass-ranges}, together with those of some 3HDMs and that of the model (R-II-1a) explored here.

The main reasons some mass regions are {\it excluded}, are qualitatively as follows:
\begin{itemize}
\item
Below $m_\text{DM}={\cal O}(50)~\text{GeV}$, it would annihilate too fast in the early Universe via off-shell Higgs bosons decaying to fermions. For $m_\text{DM} \gsim 5~\text{GeV}$ the leading contribution is from decay into $b$-quarks. Also, much of this range is excluded by LUX~\cite{Akerib:2016vxi}, by PandaX-II~\cite{Cui:2017nnn}, and by XENON1T~\cite{Aprile:2018dbl} data.

\item In the case of small mass splitting, resonant coannihilation into gauge bosons occurs for 40--45~GeV. Another resonant coannihilation, mass splitting independent, at ${m_\text{DM} \sim 60~\text{GeV}}$ is due to annihilation into the Higgs boson, dependent on $\lambda_L$.

\item
In the region between ${\cal O}(80)~\text{GeV}$ and ${\cal O}(m_h)$ it annihilates too fast via a pair of gauge or Higgs bosons. The latter channel, however, depends on $\lambda_L$.

\item
In the region from ${\cal O}(m_h)$ to ${\cal O}(500)~\text{GeV}$ it annihilates too fast via two on- or off-shell gauge bosons. Depending on the $\lambda_L$ value, annihilation into a pair of Higgs bosons can become significant. Also, in this region, annihilation into $t$-quarks becomes available.

\item
Above ${\cal O}(500)~\text{GeV}$ an important mechanism is due to the coupling to longitudinal vector bosons \cite{Hambye:2009pw}. With an increasing DM mass the relic density would grow. At some tens of TeV, interactions would become non-perturbative as a large value of $\lambda_L$ would be required to match the DM relic density value. Coannihilations between all scalars from the inert doublet become important in the limit of degeneracy.
\end{itemize}
The exact mass values where these phenomena set in, depend on details of the model and the adopted experimental constraints.
\subsection{3HDMs}
In models with three scalar doublets one has more flexibility in accommodating dark matter: 
\begin{enumerate}
\item
By having two non-inert doublets along with one inert doublet \cite{Grzadkowski:2009bt,Grzadkowski:2010au,Osland:2013sla,Merchand:2019bod}.
\item
By having one non-inert doublet along with two inert doublets.
\end{enumerate}

This latter approach has been pursued by various groups \cite{Machado:2012ed,Keus:2013hya,Fortes:2014dca,Keus:2014jha,Aranda:2014lna,Keus:2015xya,Cordero-Cid:2016krd,Cordero:2017owj,Aranda:2019vda}. The models studied in Refs.~\cite{Machado:2012ed,Fortes:2014dca} assume an $S_3$-symmetric 3HDM potential, and take the $S_3$ singlet to represent the SM-like active doublet. There are then two inert doublets with zero vacuum expectation value (vev), corresponding to R-I-1 in our terminology in table~\ref{Table:S3-vacua} below. This model has three mass degeneracies: in the charged sector as well as in the CP-odd and CP-even neutral sectors of the $S_3$ doublet, as discussed in detail in Ref.~\cite{Kuncinas:2020wrn}.
In all these models, like in the original IDM, the lightest neutral scalar in the inert sector is prevented from decaying to SM particles by a $\mathbb{Z}_2$ symmetry. In Refs.~\cite{Machado:2012ed,Fortes:2014dca} the $\mathbb{Z}_2$ symmetry is softly broken, in order to lift the degeneracy, leading to consistent models with mass, at least, in the range 40--150~GeV \cite{Fortes:2014dca}.

In Refs.~\cite{Keus:2014jha,Keus:2015xya,Cordero:2017owj}, a $\mathbb{Z}_2$-symmetric potential is constructed, and again the vacuum with two vanishing vevs is studied, yielding a model with two inert doublets. The authors point out that having more inert fields allows for more coannihilation channels, and opens up the parameter space compared to the IDM. Consistent models are found with mass in the range 53 to 77~GeV. Also, it was found that the high-mass region, which for the IDM requires a mass above some 500--525~GeV,  can give consistent models with mass down to 360~GeV \cite{Keus:2015xya}. 

In Ref.~\cite{Cordero-Cid:2016krd}, the $\mathbb{Z}_2$-symmetric 3HDM is studied, with two inert doublets and a new ingredient being explicit CP violation via complex coefficients in the $\mathbb{Z}_2$-symmetric potential. This allows for more parameters, but only a modest extension of the mass ranges allowed by the CP-conserving 3HDM discussed above is found.

Finally, one can stabilise DM by a $\mathbb{Z}_3$ symmetry, rather than the $\mathbb{Z}_2$ symmetry~\cite{Aranda:2014lna,Aranda:2019vda}. In this case, two-component DM is considered.

\section{The \boldmath$S_3$-symmetric models}
\label{sect:S3-potential}
\setcounter{equation}{0}

Our philosophy here is to first identify vacua with at least one vanishing vev as possible frameworks for a DM model. We shall work in the irreducible representation, where we have two SU(2) doublets, $h_1$ and $h_2$ in a doublet of $S_3$, and one $S_3$ singlet denoted $h_S$. The Higgs doublets with a vanishing vev are labelled as ``inert'', since their lightest member (assumed neutral) could be DM. They are listed in table~\ref{Table:S3-vacua} below. Some of these are stabilised by a surviving symmetry of the potential, whereas others would require an imposed $\mathbb{Z}_2$ symmetry. The general $S_3$-symmetric scalar potential is invariant under $\mathbb{Z}_2:\, h_1 \to -h_1$. We do not consider other mechanisms which would stabilise the DM candidate.

Some of these vacua are associated with massless states, hence we shall allow for soft breaking of the $S_3$ symmetry in the potential \cite{Kuncinas:2020wrn}, noting that soft breaking is not possible in the Yukawa sector.
When introducing soft breaking terms, the constraints will change. However, we will retain the nomenclature of the unbroken case from which they originate, thus when adding soft-breaking terms to R-I-1, we denote it r-I-1.

\subsection{The scalar potential}

In terms of the $S_3$ singlet ($\mathbf{1}:h_S$) and doublet ($\mathbf{2}:\left(h_1\,h_2\right)^\mathrm{T}$) fields, the $S_3$-symmetric potential can be written as \cite{Kubo:2004ps,Teshima:2012cg,Das:2014fea}:
\begin{subequations} \label{Eq:V-DasDey}
\begin{align}
V_2&=\mu_0^2 h_S^\dagger h_S +\mu_1^2(h_1^\dagger h_1 + h_2^\dagger h_2), \\
V_4&=
\lambda_1(h_1^\dagger h_1 + h_2^\dagger h_2)^2 
+\lambda_2(h_1^\dagger h_2 - h_2^\dagger h_1)^2
+\lambda_3[(h_1^\dagger h_1 - h_2^\dagger h_2)^2+(h_1^\dagger h_2 + h_2^\dagger h_1)^2]
\nonumber \\
&+ \lambda_4[(h_S^\dagger h_1)(h_1^\dagger h_2+h_2^\dagger h_1)
+(h_S^\dagger h_2)(h_1^\dagger h_1-h_2^\dagger h_2)+\hc] 
+\lambda_5(h_S^\dagger h_S)(h_1^\dagger h_1 + h_2^\dagger h_2) \nonumber \\
&+\lambda_6[(h_S^\dagger h_1)(h_1^\dagger h_S)+(h_S^\dagger h_2)(h_2^\dagger h_S)] 
+\lambda_7[(h_S^\dagger h_1)(h_S^\dagger h_1) + (h_S^\dagger h_2)(h_S^\dagger h_2) +\hc]
\nonumber \\
&+\lambda_8(h_S^\dagger h_S)^2.
\label{Eq:V-DasDey-quartic}
\end{align}
\end{subequations}
Note that there are two coefficients in the potential that could be complex, thus CP can be broken explicitly. For simplicity, we have chosen all coefficients to be real. In spite of this choice there remains the possibility of breaking CP spontaneously.

In the irreducible representation, the $S_3$ doublet and singlet fields will be decomposed as
\begin{equation} \label{Eq:hi_hS}
h_i=\left(
\begin{array}{c}h_i^+\\ (w_i+\eta_i+i \chi_i)/\sqrt{2}
\end{array}\right), \quad i=1,2, \qquad
h_S=\left(
\begin{array}{c}h_S^+\\ (w_S+ \eta_S+i \chi_S)/\sqrt{2}
\end{array}\right),
\end{equation}
where the $w_i$ and $w_S$ parameters can be complex.

The symmetry of the potential can be softly broken by the following terms \cite{Kuncinas:2020wrn}:
\begin{equation}\label{VSoftGenericBasis}
\begin{aligned}
V_2^\prime =&\,\mu_2^2 \left( h_1^\dagger h_1 - h_2^\dagger h_2 \right) 
+ \frac{1}{2} \nu_{12}^2 \left( h_1^\dagger h_2 + \mathrm{h.c.} \right)\\&
+ \frac{1}{2} \nu_{01}^2 \left( h_S^\dagger h_1 + \mathrm{h.c.}\right)+ \frac{1}{2} \nu_{02}^2 \left( h_S^\dagger h_2 + \mathrm{h.c.} \right).
\end{aligned}
\end{equation}
In accordance with the previous simplification of couplings we assume that the soft terms are real. Soft symmetry-breaking terms are required whenever we work with solutions with $\lambda_4=0$ since in this case most vacua lead to massless scalar states, Goldstone bosons of an O(2) symmetry resulting from this choice. In our analysis we shall not make use of these terms.

We recall that in order for a doublet to accommodate a DM candidate it must have a vanishing vev, since otherwise it would decay via its gauge couplings, e.g., the $SW^+W^-$ and $SZZ$ couplings.

\subsection{The Yukawa interaction}

Whenever the singlet vev, $w_S$, is different from zero we can construct a trivial Yukawa sector, ${\mathcal{L}_Y \sim 1_f \otimes 1_h}$. In this case, the fermion mass matrices are:
\begin{subequations}\label{Eq.FMMws}
\begin{align}
{\cal M}_u=\frac{1}{\sqrt{2}} \diag\left(y_1^u,\,y_2^u,\,y_3^u\right)w_S^\ast,\\
{\cal M}_d=\frac{1}{\sqrt{2}} \diag\left(y_1^d,\,y_2^d,\,y_3^d\right)w_S,
\end{align}
\end{subequations}
where $y$'s are the Yukawa couplings of appropriate fermions.

Another possibility is when fermions transform non-trivially under $S_3$, with a Yukawa Lagrangian written schematically as ${\mathcal{L}_Y \sim (2 \oplus 1)_f \otimes (2 \oplus 1)_h}$, one doublet and one singlet of $S_3$, 
\begin{equation*}
\mathbf{2}:\left(Q_1\,Q_2\right)^\mathrm{T},\,\left(u_{1R}\,u_{2R}\right)^\mathrm{T},\,\left(d_{1R}\,d_{2R}\right)^\mathrm{T}\quad\text{and}\quad\mathbf{1}:Q_3,\,u_{3R},\,d_{3R}.
\end{equation*}
Such structure yields the mass matrix for each quark sector ($d$ and $u$) of the form
\begin{subequations} \label{Eq:M}
\begin{align}
{\cal M}_u=
\frac{1}{\sqrt{2}}\begin{pmatrix}
y_1^uw_S^\ast+y_2^u w_2^\ast & y_2^u w_1^\ast & y_4^u w_1^\ast \\
y_2^u w_1^\ast & y_1^u w_S^\ast-y_2^u w_2^\ast & y_4^u w_2^\ast \\
y_5^u w_1^\ast & y_5^u w_2^\ast & y_3^u w_S^\ast
\end{pmatrix},\\
{\cal M}_d=
\frac{1}{\sqrt{2}}\begin{pmatrix}
y_1^dw_S+y_2^dw_2 & y_2^d w_1 & y_4^d w_1 \\
y_2^d w_1 & y_1^d w_S-y_2^d w_2 & y_4^d w_2 \\
y_5^d w_1 & y_5^d w_2 & y_3^d w_S
\end{pmatrix}.
\end{align}
\end{subequations}
If, for simplicity we assume all $y$'s to be real, then a complex Cabibbo--Kobayashi--Maskawa (CKM) matrix would have to be generated from complex vacua. As we point out in the sequel, in some cases realistic quark masses and mixing can only be generated if the quarks are taken to be $S_3$ singlets and only couple to $h_S$, which requires complex Yukawa couplings as in the SM. On the other hand, even in some of the cases when the vevs are complex, the CKM matrix is not always complex.

The hermitian quantity,
\begin{equation}
{\cal H}_f={\cal M}_f {\cal M}_f^\dagger,
\end{equation}
is going to be useful in our discussion. The fermion mass matrices $\mathcal{M}_{f}$ are diagonalised in terms of the left-handed and the right-handed fermion rotation matrices, which, in general, are not equal. However, the quantity ${\cal M}_f {\cal M}_f^\dagger$ is diagonalised in terms of the left-handed rotation matrix only, and ${\cal M}_f^\dagger {\cal M}_f$ accordingly in terms of the right-handed rotation matrix. The eigenvalues of ${\cal H}_f$ will be the squared fermion masses. 

It is instructive to count the number of parameters available in these models, and compare with the number of physical quantities to be fitted. The full Yukawa sector, eq.~(\ref{Eq:M}), has 10 parameters (5 $y_i^d$ and 5 $y_i^u$). If $w_S=0$, this number is reduced to 6 ($y_1^{(u,\,d)}=y_3^{(u,\,d)}=0$). If either $w_1=0$ or $w_2=0$, we still have 10 parameters, but if both are zero, $w_1=w_2=0$, the number is reduced to 4. The available parameters will be ``used'' to fit 6 quark masses plus 4 parameters of the CKM matrix. Apart from that, the most general vacuum configuration is given by 3 absolute values and 2 complex phases. 

As we are interested in a DM candidate, one of the requirements is to have a field with a vanishing vacuum expectation value. Let us briefly consider what happens with the Yukawa sector in this case. When the DM candidate resides in the scalar $S_3$ singlet, $w_S = 0$, we need the fermions to couple to the $S_3$ doublet, schematically represented by ${\mathcal{L}_Y \sim 2_f \otimes 2_h}$. In some cases, the DM candidate can reside in the scalar $S_3$ doublet. To keep the notation simple, we still write the Yukawa sector as ${\mathcal{L}_Y \sim (2 \oplus 1)_f \otimes (2 \oplus 1)_h}$, as the general form of the fermion mass matrices persists. However, in order to stabilise the DM candidate one needs to introduce an additional $\mathbb{Z}_2$ symmetry in the Yukawa sector to decouple a specific inert doublet from the fermionic sector.

\subsection{Vacua with at least one vanishing vev}
\label{sect:vev-zero}

In table~\ref{Table:S3-vacua}, we list the vacua of the $S_3$-symmetric potential that could accommodate DM \cite{Emmanuel-Costa:2016vej}. The vacua will be represented in the form
\begin{equation}
\left(w_1,w_2,w_S\right).
\end{equation}
Whenever $\langle h_S^0\rangle\neq0$ we indicate that the fermions may transform trivially under $S_3$. This is the simplest choice. In all other cases we need them to transform non-trivially in order to acquire masses. Whenever $\lambda_4=0$, the scalar potential acquires an additional O(2) symmetry between $h_1$ and $h_2$, which could be spontaneously broken. This breaking leads to one massless neutral scalar, see Ref.~\cite{Kuncinas:2020wrn} for a classification of massless states. Another interesting feature is that the scalar potential with the $\lambda_4=0$ constraint will have two additional $\mathbb{Z}_2$ symmetries beyond $h_1\to-h_1$, involving $h_2$ and $h_S$. 
The symmetry $h_1\to-h_1$ is a symmetry of the potential in the irreducible representation. If $h_1$ does not acquire a vev it remains as an unbroken symmetry. Column~2 lists the vevs in the irreducible representation. $S_2$ and $S_3$ symmetry in the fourth column refer to remnant symmetries explicit in the defining representation, as does ``cyclic $\mathbb{Z}_3$''. For $\lambda_4\neq0$ real vacua can at most break $S_3\to S_2$ \cite{Derman:1979nf}. Whenever $\lambda_4=0$ the $S_3$ symmetry can be fully broken by real vacua.

{\renewcommand{\arraystretch}{1.24}
\begin{table}[htb]\footnotesize
\caption{$S_3$ vacua that might accommodate DM due to a vanishing vev~\cite{Emmanuel-Costa:2016vej}. The ``hat'', $\hat w_i$, denotes an absolute value. See the text for further explanations.}
\label{Table:S3-vacua}
\begin{center}
\begin{tabular}{|c|c|c|c|c|c|c|c|}
\hline\hline
Vacuum & vevs & $\lambda_4$ & symmetry & \begin{tabular}[c]{@{}c@{}} \# massless \\ states \end{tabular} & fermions under $S_3$ \\
\hline
\hline
R-I-1 & $(0,0,w_S)$ & $\surd$ & $S_3$, $h_1\to -h_1$ & none & trivial \\
\hline
R-I-2a & $(w,0,0)$ & $\surd$  & $S_2$ & none & non-trivial \\
\hline
R-I-2b,2c & $(w,\pm\sqrt3w,0)$ & $\surd$  & $S_2$ & none & non-trivial \\
\hline
R-II-1a & $(0,w_2,w_S)$  & $\surd$ & $S_2$, $h_1\to-h_1$ & none & trivial \\
\hline
R-II-2 & $(0,w,0)$ & $0$ & $S_2$, $h_1\to -h_1, h_S\to -h_S$ & 1 & non-trivial \\
\hline
R-II-3 & $(w_1,w_2,0)$ & $0$ & $h_S\to-h_S$ & 1 & non-trivial \\
\hline
R-III-s & $(w_1,0,w_S)$ & $0$ & $h_2\to-h_2$ & 1 & trivial \\
\hline
C-I-a & $(\hat w_1,\pm i \hat w_1,0)$ & $\surd$ & cyclic $\mathbb{Z}_3$ & none & non-trivial \\
\hline
C-III-a & $(0, \hat w_2e^{i\sigma_2},\hat w_S)$ & $\surd$ & $S_2$, $h_1\to-h_1$ & none & trivial \\
\hline
C-III-b & $(\pm i \hat w_1,0,\hat w_S)$ & 0 & $h_2\to-h_2$ & 1 & trivial \\
\hline
C-III-c & $(\hat w_1e^{i\sigma_1},\hat w_2e^{i\sigma_2},0)$ & 0 & $h_S\to-h_S$ & 2 & non-trivial \\
\hline
C-IV-a & $(\hat w_1 e^{i\sigma_1},0,\hat w_S)$ & 0 & $h_2\to-h_2$ & 2 & trivial \\
\hline
\hline
\end{tabular}
\end{center}
\end{table}}

Below, we indicate some pros and cons of different vacua of $S_3$-3HDM, following the nomenclature of Ref.~\cite{Emmanuel-Costa:2016vej}.

\begin{itemize}
\item
\textbf{R-I-1: \boldmath$(0,\,0,\,w_S)$}\\
This case might result in a viable DM candidate. A related case was studied in Refs.~\cite{Machado:2012ed,Fortes:2014dca}. In order to stabilise $h_2$ they forced $\lambda_4=0$ and found that this model may result in a viable DM candidate. 

Scalar sector: The DM candidate resides in $h_1$ and thus is automatically stabilised. There are three pairs of mass-degenerate states: a charged pair and two neutral pairs. In principle, one could lift this degeneracy by softly breaking the $S_3$ symmetry of the scalar potential. The only possible soft breaking term is $\mu_2^2$. If this term is present, there are no mass degeneracies. In addition, it is possible to stabilise the $h_2$ doublet by forcing $\lambda_4=0$. There is no spontaneous symmetry breaking associated with the inert doublets and therefore no Goldstone states would arise due to spontaneously broken O(2).

Yukawa sector: ${\mathcal{L}_Y \sim 1_f \otimes 1_h}$ can give realistic fermion masses. Due to freedom of parameters, this can give a realistic CKM matrix and no flavour-changing neutral currents (FCNC). 

\item
\textbf{R-I-2a: \boldmath$(w,\,0,\,0)$}\\
The Yukawa sector is unrealistic.

Scalar sector: The $\mathbb{Z}_2$ symmetry is preserved for ${(h_2,h_S) \to -(h_2,h_S)}$, or equivalently this translates into $h_1 \to - h_1$, and thus we have two stabilised inert doublets.

Yukawa sector: ${\mathcal{L}_Y \sim 2_f \otimes 2_h}$ results in ${\det (\mathcal{H}_f)=0}$. This indicates that one of the fermion mass eigenvalues vanishes, i.e., there will be a massless fermion.

\item
\textbf{R-I-2b,2c: \boldmath$(w,\,\pm\sqrt3w,\,0)$}\\
The $S_3$ symmetry of the scalar potential needs to be softly broken. The Yukawa sector is unrealistic.

Scalar sector: There is mixing present in the mass-squared matrix as $\lambda_4 \neq 0$ and thus DM is not stabilised. If we artificially put $\lambda_4=0$, one of the neutral states would become massless. The DM is left stabilised only if the soft breaking term $\mu_2^2$ together with $\nu_{12}^2$ are introduced.

Yukawa sector:  ${\mathcal{L}_Y \sim 2_f \otimes 2_h}$ results in ${\det (\mathcal{H}_f)=0}$.

\item
\textbf{R-II-1a: \boldmath$(0,\, w_2,\, w_S)$}\\
This case results in a viable DM candidate provided that the Yukawa sector is trivial.

Scalar sector: The DM candidate resides in $h_1$ and thus is automatically stabilised.

Yukawa sector: ${\mathcal{L}_Y \sim (2 \oplus 1)_f \otimes (2 \oplus 1)_h}$ can give realistic fermion masses. The CKM matrix splits into a block-diagonal form and thus is unrealistic. However, there is another possibility to construct the Yukawa Lagrangian of the form ${\mathcal{L}_Y \sim 1_f \otimes 1_h}$. 

\item
\textbf{R-II-2: \boldmath$(0,\,w,\,0)$}\\
The $S_3$ symmetry of the scalar potential needs to be softly broken. The Yukawa sector is unrealistic.

Scalar sector: Due to $\lambda_4=0$, $\mathbb{Z}_2$ is preserved for both $h_1$ and $h_S$. The $\lambda_4=0$ constraint results in an additional Goldstone state. The only soft breaking term which does not survive minimisation is $\nu_{12}^2$. Also, $\nu_{02}^2$ cannot be the only soft breaking term as then the massless state would survive. Then, the $\mu_2^2$ and $\nu_{01}^2$ couplings are free parameter as those do not depend on the minimisation conditions. The coupling $\nu_{02}^2$ would require $\lambda_4 \neq 0$ and therefore DM is only stabilised in  $h_1$.

Yukawa sector: ${\mathcal{L}_Y \sim 2_f \otimes 2_h}$ can give realistic masses. However, the CKM matrix is split into a block-diagonal form.

\item
\textbf{R-II-3: \boldmath$(w_1,\,w_2,\,0)$}\\
The $S_3$ symmetry of the scalar potential needs to be softly broken. The Yukawa sector is unrealistic.

Scalar sector: Due to $\lambda_4=0$, $\mathbb{Z}_2$ is preserved for $h_S$. An additional Goldstone boson is present. The possible softly broken couplings are $\mu_2^2$ and $\nu_{12}^2$. If $\mu_2^2$ is the only term present, for consistency $\nu_{12}^2=0$, this would require to impose either $w_1=0$ or $w_2=0$. When $\nu_{12}^2$ is considered, it results in $w_1 = w_2$. It is also possible to have both terms present simultaneously.

Yukawa sector: ${\mathcal{L}_Y \sim 2_f \otimes 2_h}$ can give realistic fermion mass eigenvalues. However, the CKM matrix is unrealistic and there are no free parameters to control FCNC. The Yukawa sector results in six degrees of freedom: three from the $u$-type couplings and three from the $d$-type couplings. In case of the model with both couplings, \text{r-II-3-$\mu_2^2$-$\nu_{12}^2$}, there is an additional degree of freedom in terms of the ratio between the vevs $w_1$ and $w_2$.

\item
\textbf{R-III-s: \boldmath$(w_1,\,0,\,w_S)$}\\
This case might result in a viable DM candidate provided that the $S_3$ symmetry of the scalar potential is softly broken.

Scalar sector: Due to $\lambda_4=0$, $\mathbb{Z}_2$ is preserved for $h_2$. An additional Goldstone boson is present. Possible soft symmetry breaking terms are $\mu_2^2$ and $\nu_{01}^2$. Both of these couplings are unconstrained by the potential. In Ref.~\cite{Emmanuel-Costa:2016vej} this vacuum with $w_1$ complex was denoted as C-IV-a and it was pointed out that the constraints made it real, therefore there is no need for $\lambda_7$ to be zero.

Yukawa sector: ${\mathcal{L}_Y \sim (2 \oplus 1)_f \otimes (2 \oplus 1)_h}$ can give realistic fermionic masses and the CKM matrix. However, there are FCNC present in this case. In total, there are ten Yukawa couplings and a ratio between vacuum values. Due to such high number of free parameters it might be possible to control the overall effect of FCNC. There is also a possibility to construct the Yukawa Lagrangian of the form ${\mathcal{L}_Y \sim 1_f \otimes 1_h}$. 

\item
\textbf{C-I-a: \boldmath$(\hat w_1,\,\pm i \hat w_1,\,0)$}\\
Both the scalar and Yukawa sectors are unrealistic.

Scalar sector: In order to stabilise DM in $h_S$, we are forced to impose $\lambda_4=0$. In general, this model results in two neutral mass-degenerate pairs. No soft symmetry breaking terms survive and thus there are no c-I-a models.

Yukawa sector: ${\mathcal{L}_Y \sim 2_f \otimes 2_h}$ can give realistic fermion masses. However, the CKM matrix is split into a block-diagonal form.

\item
\textbf{C-III-a: \boldmath$(0,\,\hat w_2e^{i\sigma_2},\,\hat w_S)$}\\
This case might result in a viable DM candidate provided that the Yukawa sector is trivial.

Scalar sector: The DM candidate resides in $h_1$ and thus is automatically stabilised.

Yukawa sector: ${\mathcal{L}_Y \sim (2 \oplus 1)_f \otimes (2 \oplus 1)_h}$ can give realistic fermion masses. However, the CKM matrix is split into a block-diagonal form. Another possibility is to construct the Yukawa Lagrangian of the form ${\mathcal{L}_Y \sim 1_f \otimes 1_h}$.

\item
\textbf{C-III-b: \boldmath$(\pm i \hat w_1,\,0,\,\hat w_S)$}\\
This case might result in a viable DM candidate provided that the $S_3$ symmetry of the potential is softly broken.

Scalar sector: Due to $\lambda_4=0$, DM is stabilised in $h_2$. An additional Goldstone boson is present. The only possible soft symmetry breaking term is $\mu_2^2$.

Yukawa sector: ${\mathcal{L}_Y \sim (2 \oplus 1)_f \otimes (2 \oplus 1)_h}$ results in realistic fermion masses and CKM matrix. This model has eleven free parameters. FCNC are present. Another possibility is to construct the Yukawa Lagrangian of the form ${\mathcal{L}_Y \sim 1_f \otimes 1_h}$.

\item
\textbf{C-III-c: \boldmath$(\hat w_1e^{i\sigma_1},\,\hat w_2e^{i\sigma_2},\,0)$}\\
The $S_3$ symmetry of the scalar potential needs to be softly broken. The Yukawa sector is most likely unrealistic.

Scalar sector: Due to $\lambda_4=0$, DM is stabilised in $h_S$. There are two massless states present in this model. Possible soft symmetry breaking terms are $\mu_2^2$ and $\nu_{12}^2$. Based on the soft breaking terms, vevs are altered: c-III-c-$\mu_2^2$ results in $(\pm i \hat{w}_1, \hat{w}_2, 0)$, and c-III-c-$\nu_{12}^2$ in ${(\hat{w} e^{i \sigma /2}, \hat{w} e^{-i \sigma /2}, 0)}$, whereas the presence of both terms results in $(\hat{w}_1e^{i\sigma_1},\hat{w}_2e^{i\sigma_2},0)$. For more details see~\cite{Kuncinas:2020wrn}.

Yukawa sector: ${\mathcal{L}_Y \sim 2_f \otimes 2_h}$ can give realistic fermion mass eigenvalues. When the $\mu_2^2$ or $\nu_{12}^2$ terms are considered, there are seven free parameters. Preliminary check of the model with just ${\nu_{12}^2}$ resulted in unrealistic CKM and a non-negligible FCNC contribution~\cite{KunMT}. Most likely, exactly the same situation arises when $\mu_2^2$ is added. However, the c-III-c-$\mu_2^2$-$\nu_{12}^2$ model has one additional free parameter due to unfixed vevs. Nevertheless, even if the CKM values can be fitted, there are still FCNCs that have to be controlled. 

\item
\textbf{C-IV-a: \boldmath$(\hat w_1e^{i\sigma_1},\, 0,\, \hat w_S)$}\\
This case might result in a viable DM candidate provided that the $S_3$ symmetry of the potential is softly broken.

Scalar sector: Due to $\lambda_4=0$, DM is stabilised in $h_2$. There are two massless states present in this model. Possible soft symmetry breaking terms are $\mu_2^2$ and $\nu_{01}^2$. In case of the c-IV-a-$\mu_2^2$ model, the overall phase gets fixed $(i \hat w_1, 0, \hat w_S)$.

Yukawa sector: ${\mathcal{L}_Y \sim (2 \oplus 1)_f \otimes (2 \oplus 1)_h}$ can give realistic fermion masses and CKM matrix. In total, there are twelve (eleven when only $\mu_2^2$ is present) free parameters. The FCNCs are present. Another possibility is to construct the Yukawa Lagrangian of the form ${\mathcal{L}_Y \sim 1_f \otimes 1_h}$.

\end{itemize}

All in all, there are several models with a potential DM candidate. In our classification, which is summarised below, whenever unwanted Goldstone bosons are present, we refer to the need to include soft symmetry-breaking terms. Most of the models are ruled out due to unrealistic Yukawa sector. Full analysis of the Yukawa sector is out of scope of this paper. It should be noted that the non-trivial Yukawa sector might result in non-negligible FCNC and thus some of the models would be ruled out. Whenever the quarks transform trivially under $S_3$ they can only couple to one Higgs doublet, the $S_3$ singlet, and thus FCNC are not present. We have the following possible models, indicating where the DM candidate could reside and the Yukawa sector:
\begin{itemize}

\item R-I-1/r-I-1-$\mu_2^2$: $\mathrm{DM}\sim h_1$ or $\mathrm{DM}\sim (h_1,\,h_2)$, ${\mathcal{L}_Y \sim 1_f \otimes 1_h}$;

\item R-II-1a: $\mathrm{DM}\sim h_1$, ${\mathcal{L}_Y \sim 1_f \otimes 1_h}$;

\item r-III-s-$(\mu_2^2, \nu_{01}^2)$: $\mathrm{DM}\sim h_2$, ${\mathcal{L}_Y \sim (2 \oplus 1)_f \otimes (2 \oplus 1)_h}$ or ${\mathcal{L}_Y \sim 1_f \otimes 1_h}$;

\item C-III-a: $\mathrm{DM}\sim h_1$, ${\mathcal{L}_Y \sim 1_f \otimes 1_h}$;

\item c-III-b-$\mu_2^2$: $\mathrm{DM}\sim h_2$, ${\mathcal{L}_Y \sim (2 \oplus 1)_f \otimes (2 \oplus 1)_h}$ or ${\mathcal{L}_Y \sim 1_f \otimes 1_h}$;

\item c-III-c-$(\mu_2^2, \nu_{12}^2)$: $\mathrm{DM}\sim h_S$, ${\mathcal{L}_Y \sim 2_f \otimes 2_h}$;

\item c-IV-a-$(\mu_2^2, \nu_{01}^2)$: $\mathrm{DM}\sim h_2$, ${\mathcal{L}_Y \sim (2 \oplus 1)_f \otimes (2 \oplus 1)_h}$ or ${\mathcal{L}_Y \sim 1_f \otimes 1_h}$;

\end{itemize}

Despite the variety of models presented above, there are only three models with an $S_3$-symmetric scalar potential which is not softly broken and with a realistic Yukawa sector, that could result in a viable DM candidate. These models are ${\text{R-I-1}}$ $(0,0,w_S)$, which was covered in Refs.~\cite{Machado:2012ed,Fortes:2014dca} assuming a specific limit, ${\text{R-II-1a}}$ $(0,w_2,w_S)$, and C-III-a $(0, \hat w_2e^{i\sigma_2},\hat w_S)$.  Further on, we focus on the R-II-1a model and show that it could result in a viable DM candidate. The C-III-a model, which has spontaneous CP violation, will be presented elsewhere.

\section{The R-II-1a model}
\label{sect:R-II-1a}
\setcounter{equation}{0}

\subsection{Generalities}

The R-II-1a vacuum is defined by \cite{Emmanuel-Costa:2016vej}
\begin{equation}\label{RII1aVEV}
\{0,\,w_2,\,w_S\},
\end{equation}
and the minimisation conditions are:
\begin{subequations}
\begin{align}
\mu _0^2&= \frac{1}{2}\lambda _4\frac{ w_2^3}{w_S}-\frac{1}{2} \lambda_a w_2^2-\lambda _8 w_S^2, \\
\mu _1^2&= -\left( \lambda _1+ \lambda _3\right) w_2^2+\frac{3}{2} \lambda _4 w_2 w_S-\frac{1}{2} \lambda_a w_S^2,
\end{align}
\end{subequations}
with
\begin{equation}
\lambda_a=\lambda_5+\lambda_6+2\lambda_7.
\end{equation}

The $\mathbb{Z}_2$ symmetry is preserved for:
\begin{equation}\label{eq.Z2h1}
h_1 \to - h_1, \quad \{h_2,\,h_S\} \to \pm\{h_2,\,h_S\}.
\end{equation}
Hence, the inert doublet is associated with $h_1$, as $\left\langle h_1 \right\rangle=0$.

A trivial Yukawa sector is assumed, ${\mathcal{L}_Y \sim 1_f \otimes 1_h}$, and thus the $S_3$ singlet is solely responsible for masses of fermions, making $w_S$ a reference point. Therefore, we define the Higgs-basis rotation angle as:
\begin{equation}
\tan\beta = \frac{w_2}{w_S}.
\end{equation}
After a suitable rephasing of the scalar doublets we chose $w_S>0$. With $w_2$ possibly negative, the Higgs basis rotation angle will be in the range $\beta \in [- \frac{\pi}{2},~\frac{\pi}{2}]$. Therefore, the vevs can be parameterised as:
\begin{equation}\label{eq.BetaAngleDefined}
w_2 = v \sin \beta,\quad
w_S = v \cos \beta, \quad w_2^2 + w_S^2 = v^2.
\end{equation}

The Higgs basis rotation is given by:
\begin{equation}\label{eq.Rbeta}
\begin{aligned}
\mathcal{R}_\beta = \frac{1}{v} \begin{pmatrix}
v & 0 & 0 \\
0 & w_2 & w_S \\
0 & -w_S & w_2
\end{pmatrix} = & \begin{pmatrix}
1 & 0 & 0\\
0 & \cos \left( \frac{\pi}{2} - \beta \right) & \sin \left( \frac{\pi}{2} - \beta \right) \\
0 & -\sin \left( \frac{\pi}{2} - \beta \right) & \cos \left( \frac{\pi}{2} - \beta \right)
\end{pmatrix},\\
 = & \begin{pmatrix}
1 & 0 & 0\\
0 & \sin \beta & \cos \beta \\
0 & -\cos \beta & \sin \beta
\end{pmatrix},
\end{aligned}
\end{equation}
so that
\begin{equation}
\mathcal{R}_\beta \begin{pmatrix}
0 \\
w_2  \\
w_S 
\end{pmatrix} = \begin{pmatrix}
0 \\ v \\ 0
\end{pmatrix}.
\end{equation}

\subsection{R-II-1a masses}

\subsubsection{Charged mass-squared matrix}

The charged mass-squared matrix in the $\{h_1^+,\,h_2^+,\,h_S^+ \}$ basis is given by:
\begin{equation}
\mathcal{M}^2_\mathrm{Ch}
=\begin{pmatrix}
\left(\mathcal{M}_\mathrm{Ch}^2\right)_{11} & 0 & 0 \\
0 & \left(\mathcal{M}_\mathrm{Ch}^2\right)_{22} & \left(\mathcal{M}_\mathrm{Ch}^2\right)_{23} \\
0 & \left(\mathcal{M}_\mathrm{Ch}^2\right)_{23} & \left(\mathcal{M}_\mathrm{Ch}^2\right)_{33}
\end{pmatrix},
\end{equation}
where
\begin{subequations}
\begin{align}
\left(\mathcal{M}_\mathrm{Ch}^2\right)_{11} &= -2\lambda_3w_2^2+\frac{5}{2}\lambda_4w_2w_S-\frac{1}{2}(\lambda_6+2\lambda_7)w_S^2, \\
\left(\mathcal{M}_\mathrm{Ch}^2\right)_{22} &= \frac{1}{2} w_S \left[ \lambda_4w_2-(\lambda_6+2\lambda_7)w_S \right], \\
\left(\mathcal{M}_\mathrm{Ch}^2\right)_{23} &= -\frac{1}{2} w_2 \left[\lambda_4w_2-(\lambda_6+2\lambda_7)w_S \right], \\
\left(\mathcal{M}_\mathrm{Ch}^2\right)_{33} &= \frac{1}{2} \frac{w_2^2}{w_S} \left[ \lambda_4w_2-(\lambda_6+2\lambda_7)w_S \right].
\end{align}
\end{subequations}

The charged mass-squared matrix is diagonalisable by the rotation~\eqref{eq.Rbeta}. Therefore, the mass eigenstates can be expressed as:
\begin{subequations}
\begin{align}
h^+ & = h_1^+,\\
G^+ & = \sin\beta\,h_2^+ + \cos\beta\,h_S^+,\\ 
H^+ & = -\cos\beta\,h_2^+ + \sin\beta\,h_S^+,
\end{align}
\end{subequations}
with masses:
\begin{subequations}
\begin{align}
m^2_{h^+} &= -2\lambda_3w_2^2+\frac{5}{2}\lambda_4w_2w_S-\frac{1}{2}(\lambda_6+2\lambda_7)w_S^2, \\
m^2_{H^+} &=  \frac{v^2}{2 w_S} \left[ \lambda_4 w_2 - \left( \lambda_6 + 2\lambda_7 \right) w_S \right].
\end{align}
\end{subequations}
Positivity of the squared masses requires the following constraints to be satisfied:
\begin{subequations}
\begin{align}
\lambda_4 & > \left( \lambda_6 + 2 \lambda_7 \right) \cot \beta,\\
\lambda_4 & > \frac{4}{5} \lambda_3 \tan \beta + \frac{1}{5} \left( \lambda_6 + 2 \lambda_7 \right) \cot \beta.
\end{align}
\end{subequations}

\subsubsection{Inert-sector neutral mass-squared matrix}

The mass terms of the neutral components of the $h_1$ doublet are already diagonal. The masses of the two neutral states are given by:
\begin{subequations}\label{Eq:RII1a_hS_Masses}
\begin{align}
m^2_{\eta} &= \frac{9}{2} \lambda_4 w_2 w_S,\\
m^2_{\chi} &= -2(\lambda_2+\lambda_3)w_2^2+\frac{5}{2}\lambda_4w_2w_S -2\lambda_7w_S^2.
\end{align}
\end{subequations}
Positivity of the masses squared requires the following constraints to be satisfied:
\begin{subequations}
\begin{align}
\lambda_4 &>\frac{4}{5}\left( \lambda_2 + \lambda_3 \right) \tan \beta + \frac{4}{5} \lambda_7 \cot \beta,\\
\lambda_4 \sin \beta &>0.
\end{align}
\end{subequations}

\subsubsection{Non-inert-sector neutral mass-squared matrix}
The neutral mass-squared matrix is block-diagonal in the basis $\{\eta_2, \,\eta_S,\, \chi_2, \,\chi_S\}$. Therefore the mass-squared matrix can be split into two blocks:
\begin{equation}
\mathcal{M}_\mathrm{Neutral}^2 = \mathrm{diag}\left(\mathcal{M}_{\eta-2S}^2 ,\,\mathcal{M}_{\chi-2S}^2\right),
\end{equation}
where ``2S'' refers to the mixing of $h_2$ and $h_S$.
The mass-squared matrix of the CP-odd sector is:
\begin{equation}
\mathcal{M}^2_{\chi-2S}
=\begin{pmatrix}
\left(\mathcal{M}^2_{\chi-2S}\right)_{11} & \left(\mathcal{M}^2_{\chi-2S}\right)_{12}\vspace{2pt} \\ 
\left(\mathcal{M}^2_{\chi-2S}\right)_{12} & \left(\mathcal{M}^2_{\chi-2S}\right)_{22}
\end{pmatrix},
\end{equation}
where
\begin{subequations}
\begin{align}
\left(\mathcal{M}^2_{\chi-2S}\right)_{11} &= \frac{1}{2} w_S \left( \lambda_4w_2-4\lambda_7w_S \right), \\
\left(\mathcal{M}^2_{\chi-2S}\right)_{12} &= -\frac{1}{2} w_2 \left(\lambda_4w_2-4\lambda_7w_S \right), \\
\left(\mathcal{M}^2_{\chi-2S}\right)_{22} &= \frac{w_2^2}{2w_S} \left( \lambda_4w_2-4\lambda_7w_S \right).
\end{align}
\end{subequations}
It can be diagonalised by performing the $\mathcal{R}_\beta$ rotation~\eqref{eq.Rbeta}. The two CP-odd states are:
\begin{subequations}
\begin{align}
G^0 & = \sin\beta\,\chi_2 + \cos\beta\,\chi_S,\\ 
A & = -\cos\beta\,\chi_2 + \sin\beta\,\chi_S,
\end{align}
\end{subequations}
with
\begin{equation}
m^2_A = \frac{v^2}{2w_S} \left( \lambda_4w_2-4\lambda_7w_S \right).
\end{equation}
Positivity of the squared masses requires
\begin{equation}
\lambda_4 > 4 \lambda_7 \cot \beta.
\end{equation}

The CP-even mass-squared matrix is:
\begin{equation}\label{Eq:RII1aNeutralM2eta}
\mathcal{M}^2_{\eta-2S}
=\begin{pmatrix}
\left(\mathcal{M}^2_{\eta-2S}\right)_{11} & \left(\mathcal{M}^2_{\eta-2S}\right)_{12}\vspace{2pt} \\
\left(\mathcal{M}^2_{\eta-2S}\right)_{12} & \left(\mathcal{M}^2_{\eta-2S}\right)_{22}
\end{pmatrix},
\end{equation}
where
\begin{subequations}
\begin{align}
\left(\mathcal{M}^2_{\eta-2S}\right)_{11} &= \frac{1}{2} w_2 \left[ 4 \left( \lambda_1 + \lambda_3 \right) w_2 - 3\lambda_4 w_S \right], \\
\left(\mathcal{M}^2_{\eta-2S}\right)_{12} &= -\frac{1}{2} w_2 \left[ 3\lambda_4 w_2 - 2\lambda_a w_S   \right], \\
\left(\mathcal{M}^2_{\eta-2S}\right)_{22} &= \frac{1}{2w_S} \left( \lambda_4 w_2^3 + 4\lambda_8 w_S^3 \right).
\end{align}
\end{subequations}

The mass-squared matrix is diagonalisable by
\begin{equation}
\mathcal{R}_\alpha = \begin{pmatrix}
\cos \alpha & \sin \alpha \\
-\sin \alpha & \cos \alpha
\end{pmatrix},
\end{equation}
with
\begin{equation}\label{Eq:RII1a_t2a}
\tan(2{\alpha}) =  \frac{2w_2w_S(-3\lambda_4w_2+2\lambda_a w_S)}{4(\lambda_1+\lambda_3)w_2^2w_S-\lambda_4 \left( w_2^3+3w_2w_S^2 \right)-4\lambda_8w_S^3}.
\end{equation}
The CP-even states are:
\begin{subequations}
\begin{align}
h & = \cos{\alpha}\,\eta_2 + \sin{\alpha}\,\eta_S,\\ 
H & = -\sin{\alpha}\,\eta_2 + \cos{\alpha}\,\eta_S,
\end{align}
\end{subequations}
with masses: 
\begin{subequations}
\begin{align}
m_h^2 &= \frac{1}{4 w_S^2} \left[ 4 \left( \lambda_1 + \lambda_3 \right) w_2^2 w_S^2 + \lambda_4 w_2 w_S\left( w_2^2 - 3 w_S^2 \right) + 4\lambda_8 w_S^4  - w_S \Delta \right],\\
m_H^2 &= \frac{1}{4 w_S^2} \left[ 4 \left( \lambda_1 + \lambda_3 \right) w_2^2 w_S^2 + \lambda_4 w_2 w_S\left( w_2^2 - 3 w_S^2 \right) + 4\lambda_8 w_S^4  + w_S \Delta \right],
\end{align}
\end{subequations}
where
\begin{equation}\label{Eq:RII1a_Delta_mhmH}
\begin{aligned}
\Delta^2 =&\, 16 \left( \lambda_1 + \lambda_3\right)^2 w_2^4 w_S^2-8 \left( \lambda_1 + \lambda_3\right) w_2^2 w_S \left[ \lambda_4 \left( w_2^3 + 3 w_2 w_S^2 \right) + 4 \lambda_8 w_S^3 \right]\\
& + 16  \lambda_a^2 w_2^2 w_S^4 - 48 \lambda_4  \lambda_a w_2^3 w_S^3
+ \lambda_4^2 \left( w_2^6 + 42 w_2^4 w_S^2 + 9 w_2^2 w_S^4 \right) \\
& + 8 \lambda_4 \lambda_8 w_2 w_S^3\left( w_2^2+3 w_S^2 \right) + 16 \lambda_8^2 w_S^6.
\end{aligned}
\end{equation}
We identify the lighter state, $h$, as the SM-like Higgs boson.

It should be noted that we identified the mass eigenstates of the CP-even sector without performing a rotation to the Higgs basis. Had we done that according to
\begin{equation}
\begin{pmatrix}
h_1^\mathrm{HB} \\
h_2^\mathrm{HB} \\
h_S^\mathrm{HB}
\end{pmatrix} \equiv \mathcal{R}_\beta \begin{pmatrix}
h_1 \\
h_2 \\
h_S
\end{pmatrix},
\end{equation}
the CP-even sector would have been diagonalised by the additional rotation 
\begin{equation}
\mathcal{R}_{\alpha^\prime} = \begin{pmatrix}
\cos \alpha^\prime & \sin \alpha^\prime \\
-\sin \alpha^\prime & \cos \alpha^\prime
\end{pmatrix},
\end{equation}
with
\begin{equation}\label{Eq:RII1a_AlphaPrime}
\alpha^\prime = \alpha + \beta - \frac{\pi}{2}.
\end{equation}

\subsubsection{Mass eigenstates}

In terms of the mass eigenstates, the SU(2) doublets can be written as:
\begin{subequations}\label{RII1aHMEGen}
\begin{align}
h_1 &= \begin{pmatrix}
h^+ \\
\frac{1}{\sqrt{2}} \left( \eta + i \chi \right)
\end{pmatrix},\\
h_2 &= \begin{pmatrix}
 \sin \beta\,G^+ - \cos \beta\,H^+\\
\frac{1}{\sqrt{2}} \left( \sin \beta\,v + \cos{\alpha}\,h - \sin{\alpha}\,H  + i\left( \sin\beta\,G^0 - \cos\beta\,A\right)\right)
\end{pmatrix},\\
h_S &= \begin{pmatrix}
 \cos \beta\,G^+ + \sin \beta\,H^+\\
\frac{1}{\sqrt{2}} \left( \cos\beta\,v + \sin{\alpha}\,h + \cos{\alpha}\,H  + i\left( \cos\beta\,G^0 + \sin\beta\,A\right)\right)
\end{pmatrix},
\end{align}
\end{subequations}
whereas in the Higgs basis the SU(2) doublets can be written as:
\begin{subequations}\label{RII1aHB}
\begin{align}
h_1^\mathrm{HB} &= \begin{pmatrix}
h^+ \\
\frac{1}{\sqrt{2}} \left( \eta + i \chi \right)
\end{pmatrix},\\
h_2^\mathrm{HB} &= \begin{pmatrix}
 G^+ \\
\frac{1}{\sqrt{2}} \left( v +  \sin(\alpha+\beta)\,h + \cos(\alpha+\beta)\,H  + i G^0 \right)
\end{pmatrix},\\
h_3^\mathrm{HB} &= \begin{pmatrix}
 H^+\\
\frac{1}{\sqrt{2}} \left( -\cos(\alpha+\beta)\,h + \sin(\alpha+\beta)\,H  + i A \right)
\end{pmatrix}.
\end{align}
\end{subequations}

The expressions for the squared masses can be inverted to yield the scalar potential couplings:
\begin{subequations}\label{eq.R-II-1aInvertedCouplings}
\begin{align}
\lambda_1&=\frac{v^2 \left[ 9 \left( m_{h^+}^2+\cos^2 {\alpha}\,  m_h^2+\sin^2{\alpha}\,  m_H^2 \right)-m_{\eta}^2 \right]-9 m_{H^+}^2 w_S^2}{18 v^2 w_2^2}, \\
\lambda_2&=\frac{ \left(m_{h^+}^2 - m_{\chi}^2 \right)v^2 + \left(m_A^2-m_{H^+}^2 \right)w_S^2}{2 v^2 w_2^2}, \\
\lambda_3&=\frac{ \left(4 m_ \eta ^2-9 m_{h^+}^2 \right)v^2+9 m_{H^+}^2 w_S^2}{18 v^2 w_2^2}, \\
\lambda_4&=\frac{2 m_ \eta ^2}{9 w_2 w_S}, \label{Eq:R-II-1aInvertedCouplings-lam4}\\
\lambda_5 &=\frac{2 m_{H^+}^2}{v^2}+\frac{w_2 m_ \eta ^2-\frac{9}{2} \sin (2\alpha) w_S (m_H^2-m_h^2)}{9 w_2 w_S^2}, \\
\lambda_6&= \frac{m_A^2-2 m_{H^+}^2}{v^2} + \frac{m_ \eta ^2}{9 w_S^2}, \\
\lambda_7&=\frac{1}{18} \left(\frac{m_ \eta ^2}{w_S^2}-\frac{9 m_A^2}{v^2}\right), \\
\lambda_8&=\frac{9 w_S^2 \left(\sin^2 {\alpha}\,  m_h^2+\cos^2 {\alpha}\,  m_H^2\right)-w_2^2 m_ \eta ^2}{18 w_S^4}.
\end{align}
\end{subequations}

The model is invariant under a simultaneous transformation of
\begin{equation}\label{Eq:Param_sym}
\begin{aligned}
&\beta \to - \beta, ~ \alpha \to \pi - \alpha,\\
&\lambda_4 \to - \lambda_4,\\
&\{H^\pm,\,H,\,A\} \to -\{H^\pm,\,H,\,A\},
\end{aligned}
\end{equation}
which could have been adopted, but is not, in order to reduce ranges of parameters.

\subsection{The R-II-1a couplings}\label{Sect:RII1a-couplings}

Below, we quote the gauge and Yukawa couplings of the R-II-1a model. The scalar-sector couplings are collected in appendix~\ref{App:Scalar_Couplings_RII1a}.

\subsubsection{Gauge couplings}

After substituting the doublets in terms of the mass eigenstates~\eqref{RII1aHMEGen} into the kinetic Lagrangian, the resulting terms are:
\begin{subequations}
\begin{align}\label{LVVH-RII1a}
\begin{split}
\mathcal{L}_{VVH} =& \left[ \frac{g}{2 \cos \theta_W}m_ZZ_\mu Z^\mu + g m_W W_\mu^+ W^{\mu-} \right] \left[ \sin (\alpha + \beta)h + \cos (\alpha + \beta)H \right],
 \end{split}\\
\begin{split}\label{LVHH-RII1a}
\mathcal{L}_{VHH} =& -\frac{ g}{2 \cos \theta_W}Z^\mu \left[ \eta \overset\leftrightarrow{\partial_\mu} \chi  - \cos (\alpha+\beta)h \overset\leftrightarrow{\partial_\mu} A + \sin (\alpha+\beta) H \overset\leftrightarrow{\partial_\mu} A \right]\\
& - \frac{g}{2}\bigg\{ i W_\mu^+ \left[ ih^- \overset\leftrightarrow{\partial^\mu} \chi + h^- \overset\leftrightarrow{\partial^\mu} \eta  - \cos (\alpha+\beta)H^- \overset\leftrightarrow{\partial^\mu} h \right. \\ & \hspace{60pt} \left.  
+ \sin (\alpha+\beta)H^-\overset\leftrightarrow{\partial^\mu}H + i H^- \overset\leftrightarrow{\partial^\mu} A \right] + \mathrm{h.c.} \bigg\}\\
& + \left[ i e A^\mu + \frac{i g}{2} \frac{\cos (2\theta_W)}{\cos \theta_W} Z^\mu \right] \left( h^+ \overset\leftrightarrow{\partial_\mu} h^- + H^+ \overset\leftrightarrow{\partial_\mu} H^-  \right),
\end{split}\\
\begin{split}\label{LVVHH-RII1a}
\mathcal{L}_{VVHH} =& \left[ \frac{g^2}{8 \cos^2\theta_W}Z_\mu Z^\mu + \frac{g^2}{4} W_\mu^+ W^{\mu-} \right] \left( \eta^2 + \chi^2 + h^2 + H^2 +A^2\right)\\
& + \bigg\{ \left[ \frac{e g}{2} A^\mu W_\mu^+ - \frac{g^2}{2} \frac{\sin^2\theta_W}{\cos \theta_W}Z^\mu W_\mu^+ \right] \left[ \eta h^- + i \chi h^- - \cos (\alpha+\beta)hH^- \right. \\ 
& \hspace{145pt} \left. + \sin (\alpha+\beta)HH^- + iAH^- \right] + \mathrm{h.c.} \bigg\}\\
&+ \left[ e^2 A_\mu A^\mu + e g \frac{\cos (2\theta_W)}{\cos \theta_W}A_\mu Z^\mu + \frac{g^2}{4} \frac{\cos^2(2\theta_W)}{\cos^2\theta_W}Z_\mu Z^\mu + \frac{g^2}{2} W_\mu^- W^{\mu +} \right] \\
&\times\left( h^-h^+ + H^-H^+ \right),
\end{split}
\end{align}
\end{subequations}
where we have left out couplings involving Goldstone fields.

From the interaction terms $Z Z h$ and $Z Z H$ it follows that the states $h$ and $H$ are CP-even and therefore the state $A$ is CP-odd. Provided that the $h$ scalar is associated with the SM-like Higgs boson, from the interactions $hZZ$ and $hW^\pm W^\mp$ it follows that the SM-like limit is reached for
\begin{equation}\label{Eq:RII1a_SV_SM}
\sin(\alpha+\beta)=1.
\end{equation}

\subsubsection{Yukawa couplings}

As noted earlier, there are two possibilities to construct the Yukawa Lagrangian:
\begin{equation*}
\begin{aligned}
&\mathcal{L}_Y \sim (2 \oplus 1)_f \otimes (2 \oplus 1)_h, ~~ \text{and }\\
&\mathcal{L}_Y \sim 1_f \otimes 1_h.
\end{aligned}
\end{equation*}
Although the first option can give realistic fermion masses, the CKM matrix splits into a block-diagonal form. We consider the trivial representation for fermions\footnote{In our study neutrino masses are of no particular interest.}:
\begin{equation}\label{Eq:LYRII1a}
- \mathcal{L}_Y  =  \overline Q_{i\,L}^{\,0} y_{ij}^d h_S d_{j\,R}^{\,0} + \overline Q_{i\,L}^{\,0} y_{ij}^u \tilde{h}_S u_{j\,R}^{\,0} + \text{(leptonic sector)} + \mathrm{h.c.},
\end{equation}
where the Yukawa couplings $y_{ij}^{\left(u,\,d\right)}$ are assumed to be real, as mentioned earlier, and $\tilde{h}_S$ is the charge conjugated of $h_S$, i.e., ${\tilde{h}_S=i\sigma_2 h_S ^\ast}$. The fermion mass matrices are to be transformed as the superscripts ``$0$'' on the fermion fields indicate a weak-basis field. 

When considering the trivial Yukawa sector, the CKM matrix, $V_\mathrm{CKM} = V_u^\dagger V_d$, can be easily fixed to match the experimental value. Moreover, there is no naturally, at tree-level, occurring source responsible for FCNC. The scalar-fermion couplings can be extracted from eq.~\eqref{Eq:LYRII1a} by transforming into the fermion mass-eigenstates basis and multiplying the appropriate coefficients, next to the fields, by $-i$:
\begin{subequations}\label{Eq:R-II-1a-Yukawa_Neutral}
\begin{align}
g \left( h \bar{f}f \right) & = -i \frac{m_f}{v}\frac{\sin\alpha}{\cos\beta}, \quad~
g \left( H \bar{f}f \right)  = -i \frac{m_f}{v}\frac{\cos\alpha}{\cos\beta},\\ 
g \left( A \bar{u}u \right) & = - \gamma_5 \frac{m_{u}}{v}\tan \beta, \quad
g \left( A \bar{d}d \right) = \gamma_5 \frac{m_{d}}{v}\tan \beta,
\end{align}
\end{subequations}
and for the leptonic sector, the Dirac mass terms would lead to similar relations. The SM-like limit for the scalar $h$, $g_{h\bar{f}f}^\mathrm{SM}=-i\,m_f /v$, is restored at 
\begin{equation}\label{Eq:RII1a_FS_SM}
\frac{\sin\alpha}{\cos\beta}= 1.
\end{equation}

Finally, the charged scalar-fermion couplings are:
\begin{subequations}\label{Eq:R-II-1a-Yukawa_Charged}
\begin{align}
g \left( H^+ \bar{u}_i d_j \right) & = i \frac{\sqrt{2}}{v} \tan \beta \left[ P_L m_u - P_R m_d  \right] \left( V_\mathrm{CKM} \right)_{ij},\\
g \left( H^- \bar{d}_i u_j \right) & = i \frac{\sqrt{2}}{v} \tan \beta \left[ P_R m_u - P_L m_d  \right] \left( V_\mathrm{CKM}^\dagger \right)_{ji},\\
g \left( H^+ \bar{\nu} l  \right) & = - i \frac{\sqrt{2} m_l}{v} \tan \beta P_R,\\
g \left( H^- \bar{l} \nu  \right) & = - i \frac{\sqrt{2} m_l}{v} \tan \beta P_L.
\end{align}
\end{subequations}

The structure of the Yukawa couplings is the same as that of the 2HDM, Type~I, except that our definition of $\tan\beta$ is the inverse, since the singlet vev is here taken as the reference (denominator):
\begin{equation}
\left(\tan\beta\right)_\text{R-II-1a}=\left(\frac{1}{\tan\beta}\right)_\text{2HDM, Type~I}.
\end{equation}
Note also that $\alpha$ is defined differently.

\section{Model analysis}
\label{sect:exp-constraints}

For simplicity, we introduce a generic notation for different scalars,
\begin{align*}
\varphi_i^\pm =& \{h^\pm,\, H^\pm\}, \\
\varphi_i =& \{\eta,\, \chi \}.
\end{align*}

Precise measurements of the $W^\pm$ and $Z$ widths at LEP~\cite{Schael:2013ita} forbid decays of the gauge bosons into a pair of scalars.  In our case, the lower limits on the scalar masses are set by the following constraints: ${m_{\varphi_i^\pm} > \frac{1}{2} m_Z}$, and $m_{\varphi_i} + m_{h^\pm} > m_{W^\pm}$, and $m_\eta + m_\chi > m_Z$. Usually, a conservative lower bound for the charged masses $m_{\varphi_i^\pm} \geq 80\text{ GeV}$ is adopted~\cite{Pierce:2007ut,Arbey:2017gmh}. We assume a slightly more generous lower bound of $m_{\varphi_i^\pm} \geq 70\text{ GeV}$.

The model is analysed using the following input:
\begin{itemize}
\item Mass of the SM-like Higgs is fixed at $m_h = 125.25$ GeV~\cite{Zyla:2020zbs};
\item The Higgs basis rotation angle $\beta \in [-\frac{\pi}{2},~\frac{\pi}{2}]$ and the $h\,\text{-}H$ diagonalisation angle $\alpha \in [0,~\pi]$;
\item The charged scalar masses $m_{\varphi_i^\pm} \in [0.07,~1]$ TeV;
\item The inert sector masses $m_{\varphi_i} \in [0,~1]$ TeV. Either $\eta$ or $\chi$ could be a DM candidate, whichever is lighter;
\item The active sector masses $\{m_H,~m_A\} \in [m_h,~1~\mathrm{TeV}]$;
\end{itemize}

For the numerical parameter scan, both theoretical and experimental constraints are evaluated. Based on the constraints, several cuts are defined and applied:
\begin{itemize}
\item Cut~1: perturbativity, stability, unitarity checks, LEP constraints;
\item Cut~2: SM-like gauge and Yukawa sector, electroweak precision observables and $B$ physics;
\item Cut~3: $h \to \{\mathrm{invisible},~\gamma \gamma \}$ decays, DM relic density, direct searches;
\end{itemize}
with each of the subsequent constraint being superimposed over the previous ones.

\clearpage
\subsection{Imposing theory constraints}
\label{sect:R-II-1a-theory}

\begin{figure}[h]
\begin{center}
\includegraphics[scale=0.23]{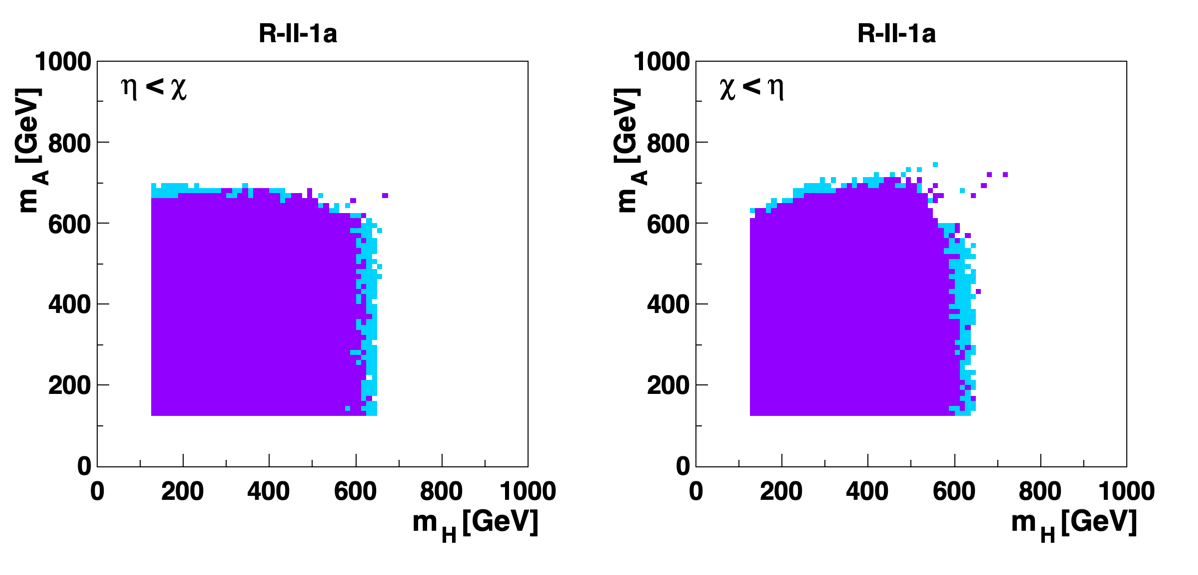} 
\includegraphics[scale=0.23]{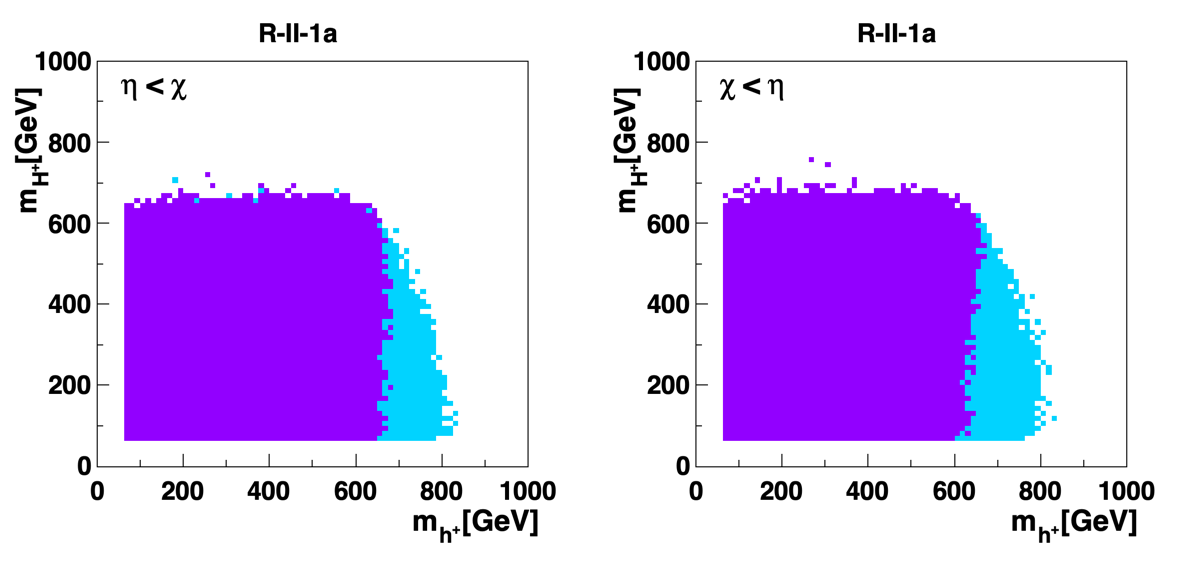} 
\includegraphics[scale=0.23]{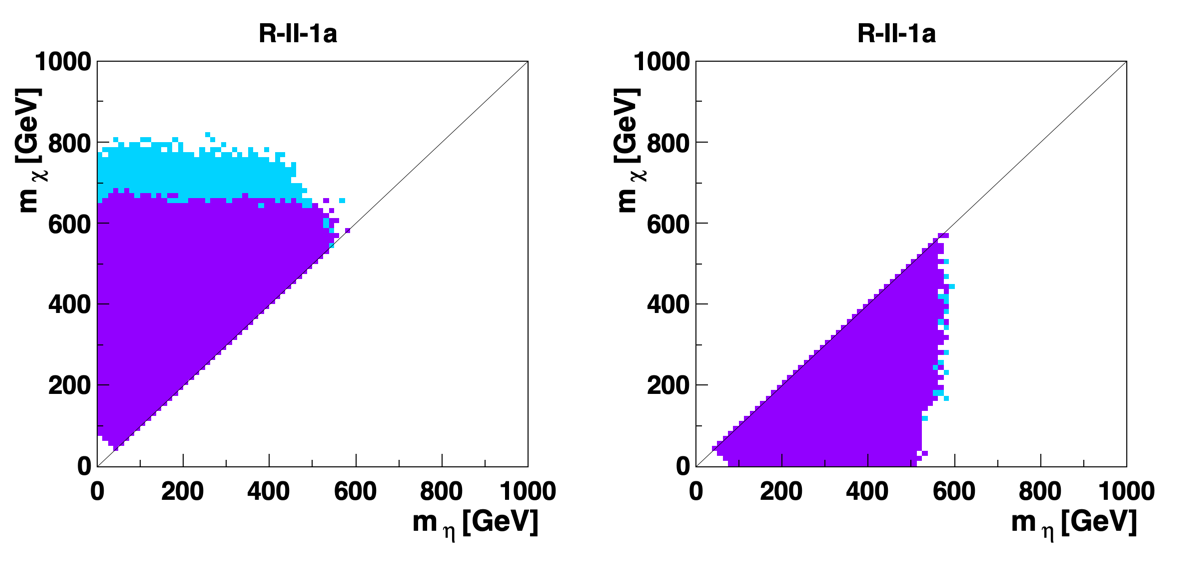}
\end{center}
\vspace*{-8mm}
\caption{Scatter plots of masses that satisfy theory constraints, Cut~1, for both orderings of $m_\eta$ and $m_\chi$.
Top: masses of the neutral states of the active doublets, $H$ and $A$.
Middle: masses of the charged states, $h^\pm$ and $H^\pm$.
Bottom: masses of the neutral states of the inert doublet, $\eta$ and $\chi$.
The light-blue region accommodates the $16\pi$ unitarity constraint, whereas the darker region satisfies the $8\pi$ constraint.}
\label{Fig:R-II-1a-masses-HA_and_charged-th_constraints}
\end{figure}

Imposing the theory constraints discussed in appendix~\ref{App:theory-constraints}, we can exclude parts of the parameter space, as illustrated in figure~\ref{Fig:R-II-1a-masses-HA_and_charged-th_constraints}. Low values of $w_2$, and hence of $\beta$ (see eq.~\eqref{eq.BetaAngleDefined}), are disfavoured by the R-II-1a model. This can be seen by inspecting the $\lambda_1$, $\lambda_2$, and $\lambda_3$ couplings~\eqref{eq.R-II-1aInvertedCouplings}, these couplings are proportional to $1/w_2^2$. A particularly instructive combination, expanded for small $\beta$, is
\begin{equation}
\lambda_1 + \lambda_3 \approx \frac{1}{6 v^2 \beta^2} \left( 3 m_h^2 + m_\eta^2 \right),
\end{equation}
with $\lambda_1$ minimised for $\alpha=\pi/2$. With a decreasing denominator $\sim \beta^{2}$, we want to control the overall value of $|\lambda_i|$, and therefore the value of the numerator must also decrease. The perturbativity constraint restricts large values of the $\lambda_i$ couplings, see appendix~\ref{App:Perturbativity}, and sets a limit $0 < \lambda_1 + \lambda _3\leq 2\pi/3$~\eqref{Eq.R_II_1a_QCLim_l1l3}. For $m_\eta^2\ll 3m_h^2$, we arrive at the bound
\begin{equation}
\frac{3 m_h^2}{4 \pi v^2} \leq \sin^2 \beta,
\end{equation}
which means that $|\tan\beta| >0.26$. For $m_\eta=m_h$ the bound is $|\tan\beta| >0.30$.

In our model fermions couple only to the $h_S$ doublet. This can be used to put a limit on the $\tan \beta$ value. From the definition of the Yukawa couplings, $Y_f = \frac{\sqrt{2} m_f}{v \cos \beta}$, and the perturbativity requirement, $|Y_f| \leq 4 \pi$, it follows that the most stringent bound comes from the heaviest state, which is $m_f = m_t$. For these values we find that $|\tan \beta| \leq 12.6$. While there would be Landau poles in the vicinity of this non-perturbative region, we note that other constraints prevent parameters from getting close.

Finally, we shall assume that the unitarity constraint is satisfied for the value of $16\pi$ rather than $8\pi$, see appendix~\ref{app:unitarity}. It turns out that points surviving all of the checks satisfy unitarity constraint at the value of $8\pi$.

\subsection{The SM-like limit}

\begin{figure}[h]
\begin{center}
\includegraphics[scale=0.35]{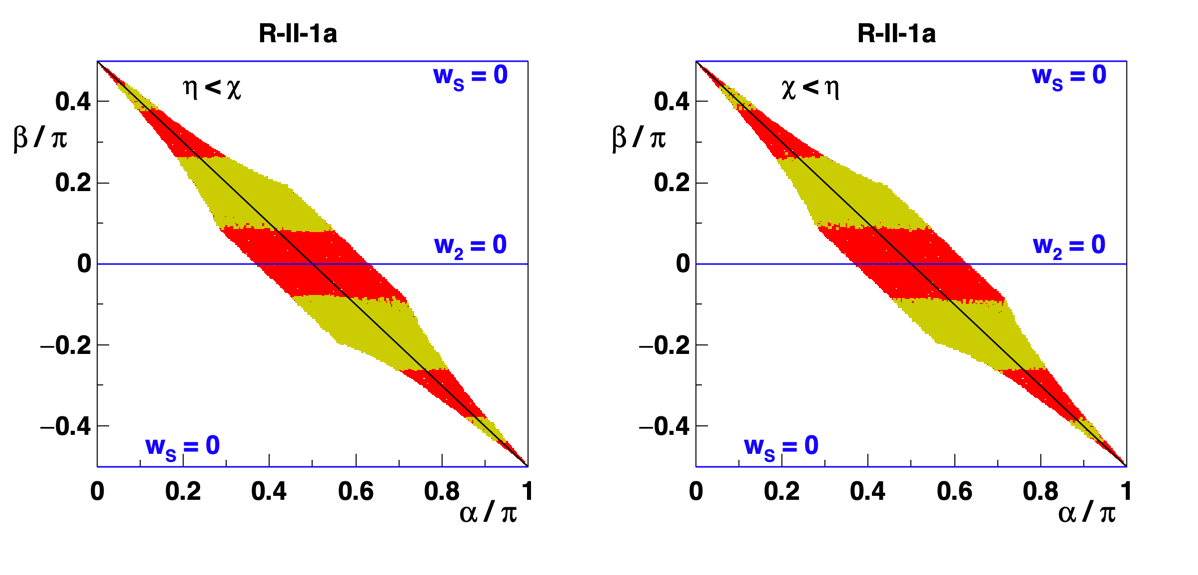}
\end{center}
\vspace*{-4mm}
\caption{Constraints on $\alpha$ and $\beta$ from the gauge and Yukawa couplings. We take $\beta \in [- \frac{\pi}{2},~\frac{\pi}{2}]$ and $\alpha \in [0,~\pi]$, see also eq.~\eqref{Eq:Param_sym}. The white region is excluded. The black diagonal line identifies the SM-like limit, which is $\alpha+\beta=\pi/2$. The coloured regions are compliant with Cut~1, whereas the yellow regions are also compliant with Cut~2 constraints. Values of $\beta$ for which $w_2$ or $w_S$ vanish are identified.}
\label{Fig:SM-Like_Limit_R-II-1a}
\end{figure}

The Standard Model like limit refers to the limit in which the CP even boson, which is in the only doublet that acquires vev when going to the Higgs basis, is already a physical boson, i.e., a mass eigenstate and therefore, does not mix with the other neutral scalars. This doublet is the one that contains the would-be Goldstone bosons $G^{\pm}$ and $G^0$. This limit is special and is referred to as the SM-like limit because this CP even  neutral boson behaves in many aspects as the SM boson. This limit is reached for $\sin (\alpha + \beta) =1$ ($\alpha =  - \beta +\pi/2$) as stated before. In this case we see from eq.~(\ref{LVVH-RII1a}) that only $h$ has couplings of the type $VVH$ and their strengths coincide with those of the SM. Furthermore, from the fact that $h$ is in the only doublet that acquires vev we see that in this limit $h$ couples to the fermions with the same strength as the SM Higgs boson. In fact, this can be seen from  eq.~(\ref{Eq:R-II-1a-Yukawa_Neutral}), since in this limit $\sin \alpha =  \cos\beta$.

We shall adopt the following 3-$\sigma$ bounds from the PDG~\cite{Zyla:2020zbs}:
\begin{subequations}\label{Eq:RII1aSMLikeLimit}
\begin{align}
\kappa^2_{VV} \equiv &\left|\sin(\alpha + \beta)\right|^2\in\{1.19\pm3\,\sigma\},\text{ which comes from $h_\mathrm{SM} W^+ W^-$,}\\
\kappa^2_{ff} \equiv &\left|\frac{\sin\alpha}{\cos\beta}\right|^2\in\{1.04\pm3\,\sigma\},\text{ which comes from $h_\mathrm{SM} \bar b b$.}
\end{align}
\end{subequations}
Note that we must impose the same sign for these two couplings of eq.~(\ref{Eq:RII1aSMLikeLimit}), in order not to spoil the interference required for $h_\text{SM}\to\gamma\gamma$.

\subsection{Electroweak precision observables}

The electroweak oblique parameters are specified by the $S$, $T$, and $U$ functions. Sufficient mass splittings of the extended electroweak sector can account for a non-negligible contribution. The $S$ and $T$ parameters get the most sizeable contributions. Results are compared against the experimental constraints provided by the PDG~\cite{Zyla:2020zbs}, assuming that $U=0$. The model-dependent rotation matrices, needed to evaluate the set of $S$ and $T$, are presented in appendix~\ref{App:Peskin_Takeuchi_Rot}.

\subsection{\boldmath$B$ physics constraints}\label{Sec:B_Physics}

The importance of a charged scalar exchange for the $\bar B\to X(s)\gamma$ rate has been known since the late 1980's \cite{Grinstein:1987pu,Hou:1988gv,Grinstein:1990tj}. The rate is determined from an expansion of the relevant Wilson coefficients in powers of $\alpha_s/(4\pi)$, starting with (1) the matching of these coefficients to the full theory at some high scale ($\mu_0\sim m_W$ or $m_t$), then (2) evolving them down to the low scale $\mu_b\sim m_b$ (in this process the operators mix), and (3) determine the matrix elements at the low scale \cite{Buras:1993xp,Ciafaloni:1997un,Ciuchini:1997xe,Borzumati:1998tg,Bobeth:1999ww,Bobeth:1999mk,Gambino:2001ew,Cheung:2003pw,Misiak:2004ew,Czakon:2006ss,Hermann:2012fc,Misiak:2015xwa,Misiak:2017bgg,Misiak:2020vlo}.

We here follow the approach of Misiak and Steinhauser \cite{Misiak:2006ab}. While the considered $S_3$-based models have two charged Higgs bosons, only one couples to fermions. This implies that we may adopt the approach used for the 2HDM with relative Yukawa couplings of the active charged scalar, eq.~(\ref{Eq:R-II-1a-Yukawa_Charged}) (in the notation of Ref.~\cite{Misiak:2006ab}),
\begin{equation} \label{Eq:Au-Ad}
A_u=A_d=\tan\beta.
\end{equation}

\begin{figure}[h]
\begin{center}
\includegraphics[scale=0.35]{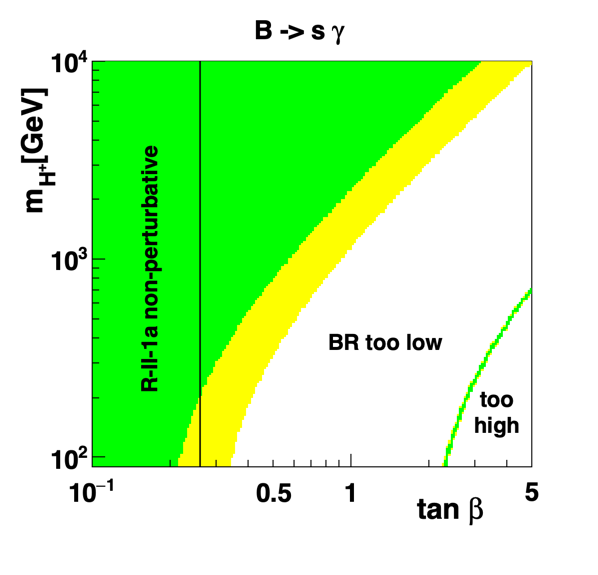}
\includegraphics[scale=0.35]{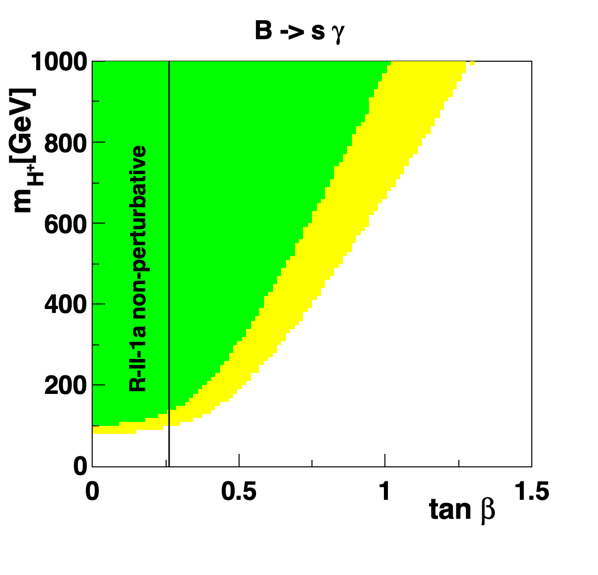}
\end{center}
\vspace*{-4mm}
\caption{Regions in the $\tan\beta$--$m_{H^+}$ plane that survive the $\bar B\to X(s)\gamma$ constraint. Left: logarithmic representation out to larger $\tan\beta$ and $m_{H^+}$. Right: Linear representation of the small-$\tan\beta$ region. The yellow region accommodates a 3-$\sigma$ tolerance with respect to the experimental rate,  whereas in the green regions, the models are within a 2-$\sigma$ bound. The vertical line at $\tan\beta=0.26$ is the lower bound on $|\tan\beta|$ compatible with $|\lambda_4|<4\pi$ for R-II-1a.}
\label{Fig:bsgamma-impact}
\end{figure}

We show in figure~\ref{Fig:bsgamma-impact} the regions in the $\tan\beta$\,--\,$m_{H^+}$ parameter plane that are not excluded by this constraint. The situation is quite different from that of the more familiar 2HDM with Type~II Yukawa couplings. According to eq.~(\ref{Eq:Au-Ad}) the relevant couplings are the same as those of the 2HDM Type~I model, with the exception that we are here interested in {\it small} values of $\tan\beta$. While the $\bar B\to X(s)\gamma$ constraint excludes large values of $\tan\beta$, low values are for R-II-1a also cut off due to the perturbativity constraint, we have $|\tan\beta|\gsim0.26$, as discussed in section~\ref{sect:R-II-1a-theory}.

Since $A_uA_d>0$, there is a region of negative interference between the SM-type contributions and the loop with the charged Higgs: as we increase the value of $\tan\beta$ for fixed $m_{H^+}$, the branching ratio will first diminish, and then at some point come back up, as illustrated in the lower right-hand corner of the left panel of figure~\ref{Fig:bsgamma-impact}. This interference region is ruled out by Cut~3. For any fixed value of $\tan\beta$, on the other hand, for sufficiently high mass $m_{H^+}$, the rate approaches the SM value.

We adopt the experimental value, $\mathrm{Br} \left( \bar B\to X(s)\gamma  \right) \times 10^4 = 3.32 \pm 0.15$  \cite{Zyla:2020zbs} and impose an $(n=3)$-$\sigma$ tolerance, together with an additional 10 per cent computational uncertainty,
\begin{equation}
\begin{aligned}
\mathrm{Br} \left( \bar B\to X(s)\gamma  \right) \times 10^4 &=  3.32 \pm \sqrt{(3.32 \times 0.1)^2 + (0.15\,n)^2}\,.
\end{aligned}
\end{equation}
The acceptable region, corresponding to the 3-$\sigma$ bound, is $[2.76;\,3.88 ]$.

\subsection{LHC Higgs constraints}

We require that the Higgs-like particle, $h$, full width is within $\Gamma_h = 3.2^{+2.8}_{-2.2}$~MeV,  which is an experimental bound adopted from \cite{Zyla:2020zbs}. In the SM the total width of the Higgs boson is around 4 MeV. The upper value, i.e., $\Gamma_h = 6~\text{MeV}$ is used in preliminary checks within the spectrum generator.

\subsubsection{Decays \boldmath$h \to \gamma \gamma $ and \boldmath$h \to g g $}

\begin{figure}[htb]
\begin{center}
\includegraphics[scale=0.35]{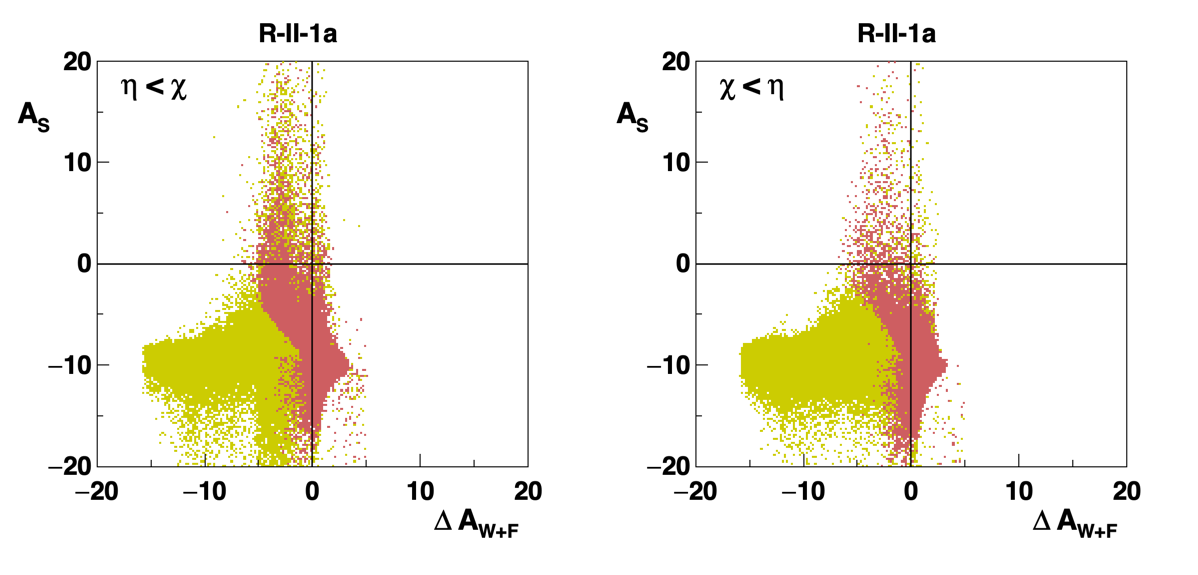}
\end{center}
\vspace*{-8mm}
\caption{Scatter plots of additional contributions to the di-photon decay amplitudes, normalised to the SM value, expressed in per cent. Complex numbers arise from the spin-dependent function $\mathcal{F}_{1/2}^S$ of equation~\eqref{Eq:F_12_S} for fermions lighter than the Higgs-like particle $h$. The $A_\mathrm{S}$ value represents the normalised contribution from the charged scalar loop, $A_\mathrm{S}/ |A_\text{SM}|$, while the $\Delta A_\text{W+F}$ value stands for an additional contribution to the SM-like part due to the $W$ and fermion loops, $\Delta A_\text{W+F}=\left(|A_\text{W+F}|-|A_\text{SM}|\right)/|A_\text{SM}|$.  Yellow dots represent parameters surviving Cut~2, while the red ones represent those surviving also LHC Higgs-particle constraints: full width, the invisible branching ratio, and the di-photon constraint.}
\label{Fig:g_hgammagamma}
\end{figure}

The di-photon partial decay width is modified by the charged-scalar loop in comparison to the SM case. The one-loop width is known~\cite{Ellis:1975ap, Shifman:1979eb, Gunion:1989we}:
\begin{equation}\label{Eq:h_gammagamma}
\begin{aligned}
\Gamma\left(h \rightarrow \gamma \gamma\right)&=\frac{\alpha^2 m_{h}^3}{256 \pi^3 v^2}\bigg|\sum_f  Q_f^2 N_c C_{\bar{f}fh}^S \mathcal{F}_{1 / 2}^S\left(\tau_{f}\right) + C_{W^+W^-h} \mathcal{F}_1\left(\tau_{W^\pm}\right) \\ 
&\hspace{70pt} + \sum_{\varphi^\pm_i}C_{ \varphi^+_i \varphi^-_i h} \mathcal{F}_0\left(\tau_{\varphi^\pm}\right) \bigg|^{2},
\end{aligned}
\end{equation}
where $\alpha$ is the fine-structure constant, $Q_f$ is the electric charge of the fermion, $N_c = 3\,(1)$ for quarks (leptons), and the $C$'s are the couplings normalised to those of the SM,
\begin{equation}
\begin{aligned}
\mathcal{L}_\text{int}^\prime  =\,&-\frac{ m_f}{v}  C_{\bar{f}fh}^S \bar{f}fh +  g m_W C_{W^+W^-h} W_\mu^+ W^{\mu -} h  - \frac{2   m_{\varphi^\pm_i}^2}{v} C_{ \varphi^+_i \varphi^-_i h}  \varphi^+_i \varphi^-_i h.
\end{aligned}
\end{equation}
The spin-dependent functions $\mathcal{F}_{1 / 2}^{S}$, $\mathcal{F}_1$, and $\mathcal{F}_0$ can be found in appendix~\ref{App:Additional}. Contributions from the fermionic loop and gauge loop versus the charged scalar loop are presented in figure~\ref{Fig:g_hgammagamma}.

In the SM case, the dominant Higgs production mechanism is through gluon fusion. However, due to experimental limitations we do not explicitly consider constraints on this channel. The rate for the two-gluon decay at the leading order is~\cite{Wilczek:1977zn,Georgi:1977gs,Ellis:1979jy,Rizzo:1979mf}
\begin{equation}
\Gamma\left(h \rightarrow g g \right) =\frac{\alpha_S^2 m_{h}^3}{128 \pi^3 v^2}
\left| \sum_f C_{\bar{f}fh}^S \mathcal{F}_{1/2}^S(\tau_f)  \right|^2,
\end{equation}
where $\alpha_S$ is the strong coupling constant. The decay width of this process can be enhanced or diminished with respect to the SM case. Such behaviour is caused by an additional factor for the amplitude, $g_{\bar{f}fh}^\text{R-II-1a} = g_{\bar{f}fh}^\text{SM}\sin\alpha/\cos\beta$.

\begin{figure}[htb]
\begin{center}
\includegraphics[scale=0.35]{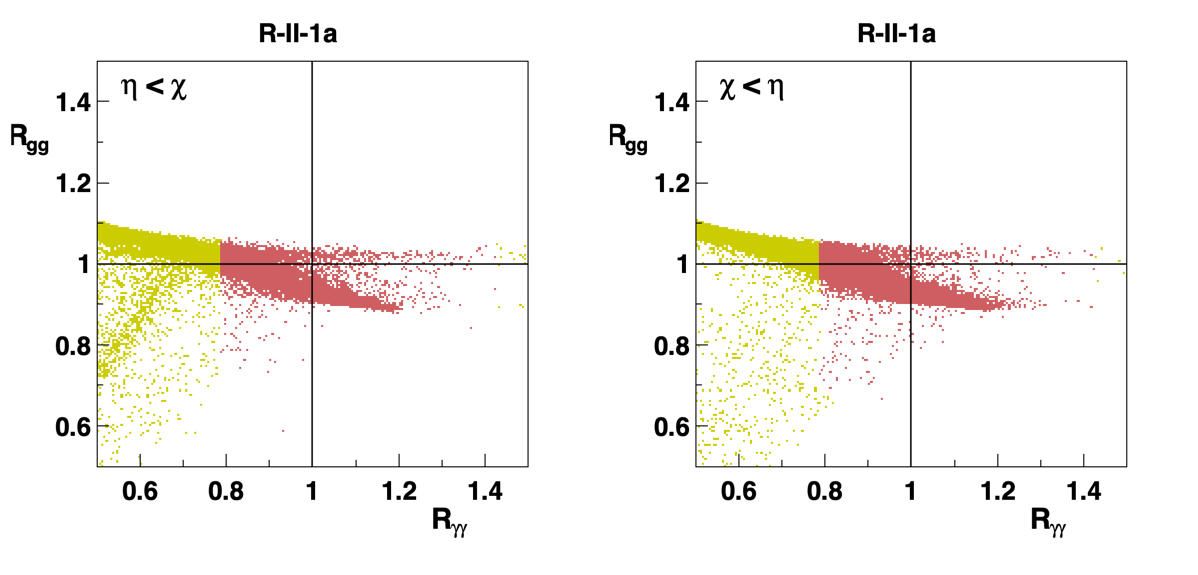}
\end{center}
\vspace*{-8mm}
\caption{Two-gluon versus di-photon Higgs-like particle branching ratios, normalised to the SM value, ${R_{ii} \equiv \text{Br}(h\to ii)/\text{Br}^\mathrm{SM}(h\to ii)}$, for different R-II-1a mass orderings. The black lines represent the SM case. A proper evaluation of the di-photon channel~\eqref{Eq:Diphoton_strength} would require an additional factor from the two-gluon channel arising from the $h$ production, i.e., each point would have to be scaled by multiplying both of the values presented in the plot, $\mu_{\gamma\gamma} = R_{gg} R_{\gamma\gamma}$. Points above the $R_{gg}=1$ line would shift right, while those below would shift left. }
\label{Fig:g_hgg}
\end{figure}

The normalised two-gluon branching ratio to the SM case is depicted in figure~\ref{Fig:g_hgg}. The gluon branching ratio for DM mass below $m_h/2$ can become low due to the opening of the invisible channel, $h\to \varphi_i\varphi_i$. However, such cases are partially excluded by other LHC Higgs-particle constraints of Cut~3. Even with more experimental data collected, the two-gluon constraint will not play a very significant role in terms of constraining the R-II-1a model. Most of the two-gluon points are within the range $\mu_{gg} \in [0.9,\,1.1]$.

In light of the above discussion, we do not aim to account for the correct $h$ two-gluon production factor and approximate the di-photon channel strength to be
\begin{equation}\label{Eq:Diphoton_strength}
\mu_{\gamma\gamma} \approx \frac{\Gamma\left( h \to \gamma \gamma \right)}{\Gamma^\text{exp}\left( h \to \gamma \gamma \right)} \frac{\Gamma^\text{exp} \left( h \right)}{\Gamma\left(h\right)},
\end{equation}
with $\mu_{\gamma \gamma} = 1.11 \pm 0.10$~\cite{Zyla:2020zbs}. We evaluate this constraint allowing for an additional 10 per cent computational uncertainty, and impose an $(n=3)\text{-}\sigma$ tolerance,
\begin{equation}\label{Eq:Diphoton_Bounds}
\mu_{\gamma\gamma} = 1.11 \pm \sqrt{(1.11\times0.1)^2 + (0.1n)^2},
\end{equation}
which corresponds to the 3-$\sigma$ range of $[0.79;\,1.43]$.

\subsubsection{Invisible decays, \boldmath $h \to\mathrm{inv.}$}

The SM-like Higgs boson can decay to the lighter scalars $h \to \varphi_i \varphi_j$ provided $2 m_{\varphi_i} < m_{h}$. If the decays are kinematically allowed, these processes can enhance the total width of the SM-like Higgs state sizeably. The observed upper limit of the invisible Higgs branching ratio reported by LHC at 95\% CL is  
\begin{subequations}
\begin{align}
\text{Br}\left( h_\mathrm{SM} \to \text{inv.} \right)&<26\%,\text{ by ATLAS~\cite{Aaboud:2019rtt}},\\
\text{Br}\left( h_\mathrm{SM} \to \text{inv.} \right)&<19\%,\text{ by CMS~\cite{Sirunyan:2018owy}}.
\end{align}
\end{subequations}

The decay width of $h$ into a pair of scalars $\varphi_i$ is given by
\begin{equation}\label{Eq.DecayWidth_SSS}
\Gamma\left( h \to \varphi_i \varphi_j \right) = \frac{2-\delta_{ij}}{32 \pi m_{h}^3} \left| g_{h \varphi_i \varphi_j} \right|^2 \sqrt{\left[ m_{h}^2 - \left( m_{\varphi_i} + m_{\varphi_j} \right)^2 \right]\left[ m_{h}^2 - \left( m_{\varphi_i} - m_{\varphi_j} \right)^2 \right]},
\end{equation}
with a symmetry factor $(2-\delta_{ij})$, where $\delta_{ij}$ is the Kronecker delta. The appropriate trilinear couplings are given by equations~(\ref{Eq.R_II_1a_hetaeta}) and (\ref{Eq.R_II_1a_hchichi}). In appendix~\ref{App:Scalar_Couplings_RII1a} the overall factor of $``-i"$  is left out\footnote{We use the notation $g_{\dots} = -i g(\dots)$.}. Due to CP conservation there is no coupling $g\left(h\,\eta\,\chi\right)$, therefore the Higgs-like particle can decay only into pairs of inert neutral scalars,
\begin{equation}
\Gamma\left( h \to \varphi_i \varphi_i \right) = \frac{1}{32 \pi m_{h}^2} \left| g_{h \varphi_i \varphi_i} \right|^2 \sqrt{ m_{h}^2 -  4m_{\varphi_i}^2}.
\end{equation}
These are the only two possible channels ($h \to \eta \eta$ and $h \to \chi \chi$) which can contribute to the invisible decay. Channels with other scalars are kinematically inaccessible due to the assumed limits, i.e., mass ordering and LEP constraints.

The experimental bounds can be applied directly if there is only a single channel open. However, there is a possibility that both inert neutral states, $\eta$ and $\chi$, can be kinematically accessible. In a simplistic approximation a particle escapes a detector of 30~meters, assuming no time dilation factor, if its lifetime exceeds a value of $\tau \gtrsim 10^{\,\text{-}7}$~seconds. The value can be expressed in terms of the total decay width, $\Gamma ^\text{tot} \lesssim 6.6 \times 10^{\,\text{-}18}$~GeV. Based on the total width of the next-to-lightest state, $\varphi_j$, two situations are possible:
\begin{itemize}
\item If the particle is not long-lived, $\Gamma^\text{tot}(\varphi_j)> 6.6 \times 10^{\,\text{-}18}$~GeV, it will decay within the detector through ${h \to \varphi_j \varphi_j \to \varphi_i \varphi_i Z^\ast Z^\ast}$, with the $Z^\ast$ subsequently also decaying. In this case only the $h \to \varphi_i \varphi_i$ channel will contribute to $\text{Br}\left( h \to \text{inv.} \right)$.
\item When $\Gamma^\text{tot}(\varphi_j)< 6.6 \times 10^{\,\text{-}18}$~GeV, the invisible branching ratio will be given by
\begin{equation}
\text{Br}\left( h \to \text{inv.} \right) = \frac{ \Gamma\left(h \to \eta \eta \right) +  \Gamma\left(h \to \chi \chi \right) }{\Gamma\left( h \right)}.
\end{equation}
\end{itemize}

In the R-II-1a model we found that the decay rate of the next-to-lightest inert neutral particle is way above $\Gamma = \mathcal{O}(10^\text{-18})$~GeV. In our calculations we shall adopt the PDG~\cite{Zyla:2020zbs} constraint, which is $\text{Br}^\mathrm{exp}\left( h \to \text{inv.} \right)<0.19$.

\subsection{The \boldmath$h$ scalar self interactions}

Let us next consider the trilinear and quadrilinear self interactions, given by eqs.~\eqref{Eq.R_II_1a_hhh} and \eqref{Eq.R_II_1a_hhhh}. The SM Higgs self-interactions are~\cite{Boudjema:1995cb}
\begin{equation}
g(h^3_\mathrm{SM})=\frac{3m_{h_\mathrm{SM}} ^2}{v},\quad
g(h_\mathrm{SM}^4)
= \frac{1}{v}g(h_\mathrm{SM}^3).
\end{equation}
In the R-II-1a model, these couplings can be expanded in terms of $m_h^2$, $m_H^2$ and $m_\eta^2$,
\begin{subequations}
\begin{align}
g(h^3)&= \frac{3 m_h^2}{v} \left[ \sin\left(\alpha+\beta\right) + \frac{2 \cos^2\left( \alpha + \beta \right)\cos\left(\alpha-\beta\right)}{\sin\left(2\beta\right)} \right] + \frac{2 m_\eta^2 \cos^3(\alpha + \beta)}{3 v \sin(2\beta)\cos^2(\beta)},\\
\begin{split}g(h^4)&= \frac{3 m_h^2}{v^2} \left[ \sin\left(\alpha+\beta\right) + \frac{2 \cos^2\left( \alpha + \beta \right)\cos\left(\alpha-\beta\right)}{\sin\left(2\beta\right)} \right]^2\\
&\hspace{15pt}+ \frac{3 m_H^2 \cos^2(\alpha + \beta) \sin^2(2\alpha)}{v^2 \sin^2(2\beta)}+ \frac{2 m_\eta^2 \cos^3(\alpha + \beta) \left[ 3 \cos(\alpha) \cot(\beta) + \sin(\alpha) \right]}{3 v^2 \sin(2\beta)\cos^3(\beta)}.\raisetag{4\normalbaselineskip}\end{split}
\end{align}
\end{subequations}
After imposing $\alpha+\beta=\pi/2$ we arrive at the SM-like Higgs couplings.

\begin{figure}[htb]
\begin{center}
\includegraphics[scale=0.35]{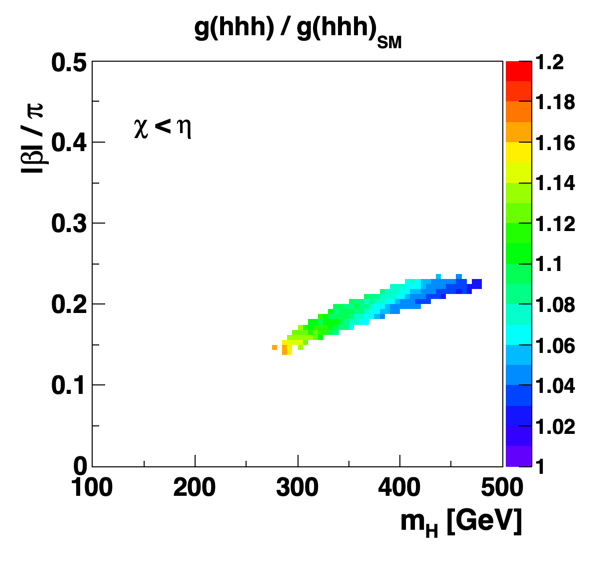}	
\includegraphics[scale=0.35]{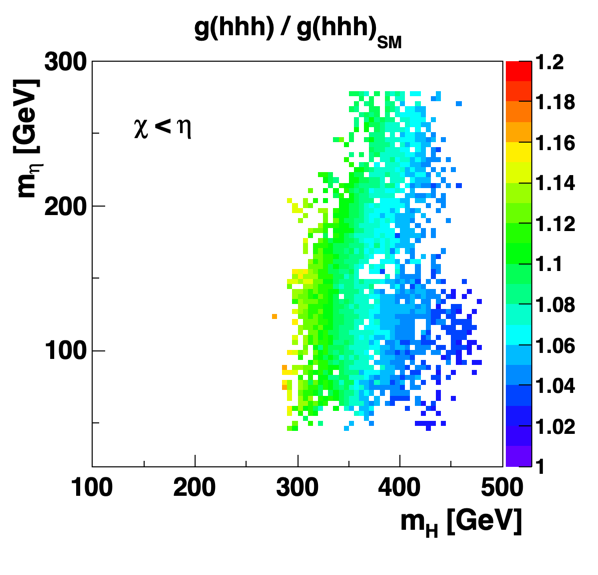}									  
\end{center}
\vspace*{-8mm}
\caption{Trilinear self interactions of the Higgs-like particle normalised to the SM value, after applying Cut~3, represented by the coloured bar. The coupling is presented as a function of the mass of the heavier CP-even state $m_H$ and the Higgs basis rotation angle $\beta$ (left) or the mass of the neutral inert scalar $m_\eta$ (right).}
\label{Fig:g_hhh_norm}
\end{figure}

In the future, the trilinear Higgs self interactions may become a crucial test for new physics. For this purpose, we show in figure~\ref{Fig:g_hhh_norm} the $h$ trilinear coupling, relative to the SM value.

\subsection{Astrophysical Observables}

We consider a standard cosmological model with a freeze-out scenario. The cold dark matter relic density along with the decay widths discussed above and other astrophysical observables are evaluated using $\mathsf{micrOMEGAs~5.2.7}$. The 't~Hooft-Feynman gauge is adopted, and switches are set to default values $\mathsf{VZdecay=VWdecay=1}$, identifying that 3-body final states will be computed for annihilation processes only. The $\mathsf{fast=-1}$ switch identifies that very accurate calculation is used. The steering $\mathsf{CalcHEP}$~\cite{Belyaev:2012qa} model files are produced with the help of $\mathsf{SARAH}$~\cite{Staub:2009bi,Staub:2013tta}.

We shall adopt the cold dark matter relic density value of $0.1200 \pm 0.0012$ taken from PDG~\cite{Zyla:2020zbs}. The relic density parameter will be evaluated using a 3-$\sigma$ tolerance and assuming an additional 10 per cent computational uncertainty,
\begin{equation} \label{Eq:Omega-value}
\begin{aligned}
\Omega h^2 &= 0.1200 \pm \sqrt{ \left( 0.1200 \times 0.1\right)^2  + (0.0012\,n)^2}\,,
\end{aligned}
\end{equation}
corrponding to the $[0.1075;\,0.1325]$ region.

{Let us recall results of figure~\ref{Fig:mass-ranges}. In the IDM, two regions compatible with the cold dark matter relic density were identified. These correspond to the high-mass region and the intermediate-mass region. The high-mass region is in agreement with the relic density due to two factors:
\begin{itemize}
\item Near-mass-degeneracy among the scalars of the inert sector. Small mass splittings correspond to tiny couplings, and different inert-scalar contributions to the annihilation will be suppressed and result in an acceptable relic density;
\item Freedom to choose the Higgs boson portal coupling $\lambda_L$. This parameter controls the trilinear $XXh$ and quartic $XXhh$ coupling, and must be sufficiently small. Here, $X$ refers to scalars of the inert sector, both charged and neutral.
\end{itemize}
In this region, the main DM annihilation is into $W^+W^-$. However, the relic density can be maintained at an acceptable level by suppressing annihilation via an intermediate $h$ boson and into a pair of $h$ bosons, illustrated in figure~\ref{Fig:PortalHMR}. The desired relic abundance can be achieved by adjusting the mass splittings.

\begin{figure}[h]
\begin{center}
\includegraphics[scale=0.75]{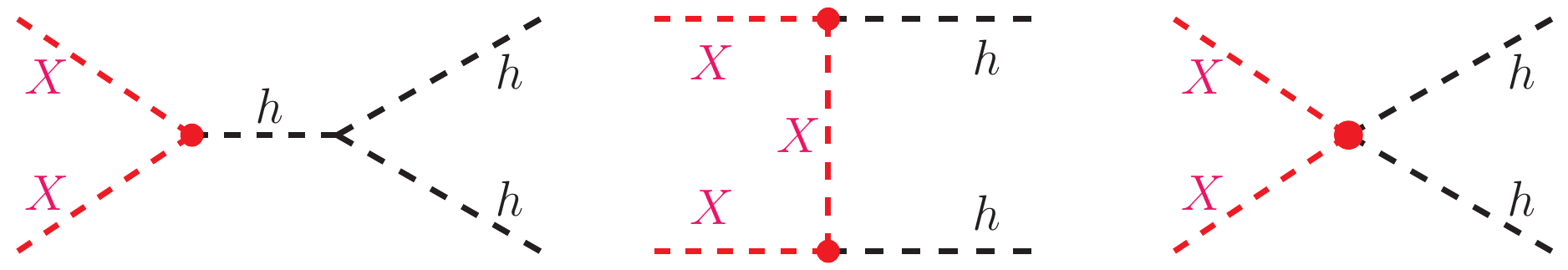}
\end{center}
\vspace*{-4mm}
\caption{Feynman diagrams contributing to $XX$ (particles of the inert sector) annihilation channels at high DM masses.}
\label{Fig:PortalHMR}
\end{figure}

Whereas the IDM and the 3HDMs considered in figure~\ref{Fig:mass-ranges} have an adjustable portal coupling (often referred to as $\lambda_L$ for the IDM), the present model is constrained by the underlying $S_3$ symmetry. Here, there is not a single portal coupling, but two: a trilinear $XXh$ and a quartic $XXhh$, plus additional ones involving other scalars. Furthermore, these are not ``free'', but correlated with other features of the model. In particular, they are constrained by the scalar masses and two angular parameters, $\alpha$ and $\beta$.

Let us consider a simplified picture with heavy active scalar bosons. At high DM masses the main DM annihilation channel is into $W^+W^-$. This channel is controlled by the gauge coupling. In addition, channels leading to $h$ scalars will be accessible, as illustrated in figure~\ref{Fig:PortalHMR}. These amplitudes are controlled by the portal couplings which should be constrained, since otherwise the DM relic density becomes too low.

To first order in $\delta$ (in the neighbourhood of the SM-like limit), where $\alpha=\pi/2-\beta+\delta$, the behaviour of the portal couplings is quite simple. In the limit of $\delta\to0$ we arrive at
\begin{equation}  \label{Eq:portal-vs-mX_sq}
\frac{g(X X h)}{v}=g(X X h h) =\frac{1}{v^2}\left[ m_h^2 + 2 m_X^2 \right].
\end{equation}
This relation shows that the portal couplings will grow with increasing DM mass.

Values of the trilinear and quartic couplings are shown in figure~\ref{Fig:quartic-coupling}. As shown in this figure, the correlation with DM mass~(\ref{Eq:portal-vs-mX_sq}) is qualitatively borne out by the parameter points surviving Cut~2. 

In the aforementioned simplification we argued that there are no good DM candidates for high mass values. In the full R-II-1a model other annihilation channels involving active scalars are also accessible. For example, there is a contribution from the $H$ scalar to the $X\,X \to h\,h$ process through the $s$-channel. Furthermore, there are other accessible active-inert scalar channels. This explains why the DM relic density is saturated at $\Omega h^2 = \mathcal{O}(10^{-4})$ at high DM values, see figure~\ref{Fig:Omega}.

\begin{figure}[htb]
\begin{center}
\includegraphics[scale=0.29]{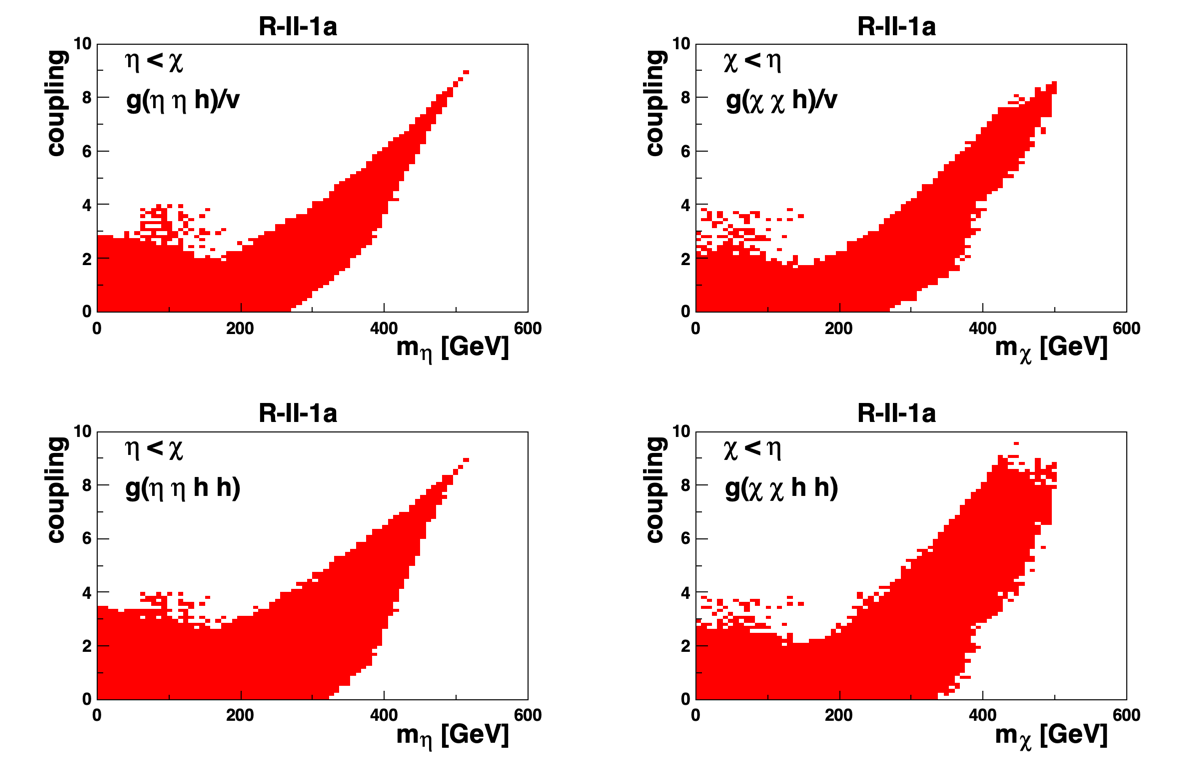}
\end{center}
\vspace*{-4mm}
\caption{Absolute value of the trilinear portal coupling $|g(XXh)/v|$ (top) and the quartic portal coupling $|g(XXhh)|$ (bottom) versus the lightest inert particle mass.}
\label{Fig:quartic-coupling}
\end{figure}

\begin{figure}[h]
\begin{center}
\includegraphics[scale=0.29]{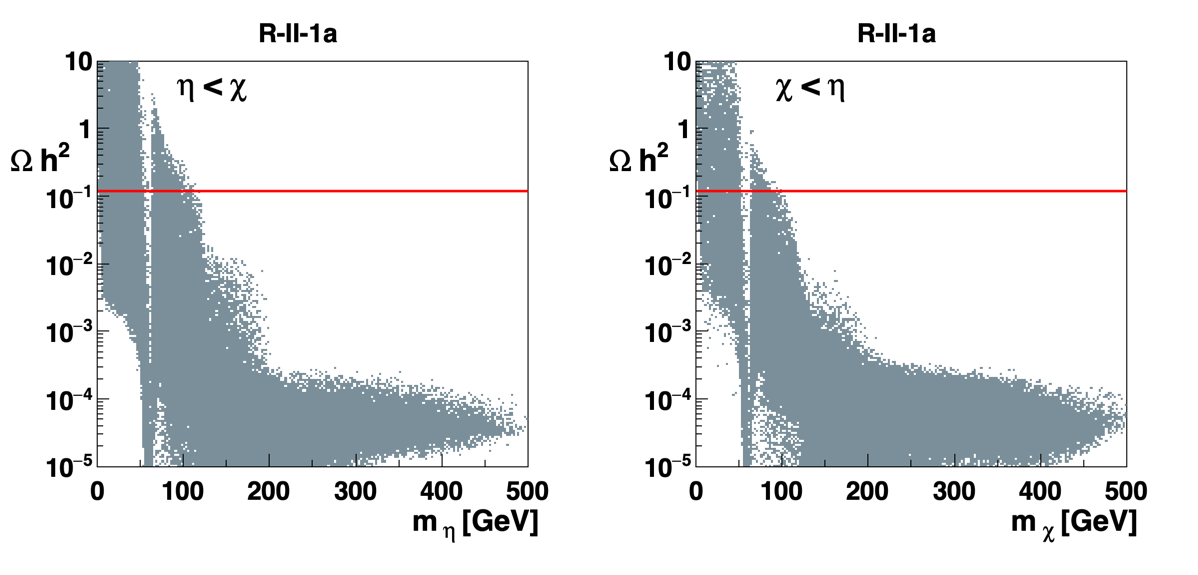}
\end{center}
\vspace*{-4mm}
\caption{Dark matter relic density for the R-II-1a model. The region compatible with the observed DM relic density does not allow for masses above around 120~GeV. In the high-mass region, ${m_{\varphi_i}\gsim500~\text{GeV}}$, the DM relic density is shown to be too low. }
\label{Fig:Omega}
\end{figure}

Going down in the DM mass, the situation changes as follows. At around $m_{\varphi_i} \approx 200$~GeV there is a kink and the maximal relic density increases from $\Omega h^2 = \mathcal{O}(10^{-4})$ up to $\Omega h^2 = \mathcal{O}(10^{-2})$. In this range, the main annihilation  (or loss) mechanisms are via the channels $\varphi_i\varphi_i\to hh$ and $\varphi_i\varphi_i\to W^\pm W^\mp$. In this same mass region many parameter points also yield $\Omega h^2 < \mathcal{O}(10^{-4})$. This happens when the dominant annihilation channels are $\varphi_i\varphi_i\to A Z$ and $\varphi_i\varphi_i\to H^\pm W^\mp$, through either $h$ or $H$ in the $s$-channel, or via $\varphi_j$ or $h^\pm$ (based on quantum numbers) in the $t$-channel.

Then, in the mass region $m_{\varphi_i} \in [m_h/2,\,120~\text{GeV}]$, the relic density ranges from above unity down to below $\Omega h^2 = \mathcal{O}\left( 10 ^{-5} \right)$. The main annihilation channels are into a pair of (virtual) $W^\pm$ bosons or $b$-quarks, with the latter becoming increasingly important as the DM mass decreases. However, there are also cases when the dominant annihilation channel is $\varphi_i\varphi_i\to gg$, which can contribute more than 50\%.

Finally, in the DM mass region corresponding to values below $m_h/2$, the primary annihilation or loss mechanism is $\varphi_i\varphi_i\to b\bar b$ trough a virtual $h$. This channel depends critically on the portal, i.e., the trilinear coupling $g(\varphi_i\varphi_ih)$.

\section{Cut 3 discussion}\label{sect:Cut_3}

It is convenient to discuss the low-mass DM situation in terms of the following four critical constraints:
\begin{itemize}
\item \textit{a} ($\Omega h^2$): DM relic density, eq.~\eqref{Eq:Omega-value};
\item \textit{b} (LHC): $h$ invisible branching ratio and $\Gamma_h \leq 6$~MeV;
\item \textit{c} (LHC): $h$  di-photon rate, eq.~\eqref{Eq:Diphoton_Bounds};
\item \textit{d} (DD): DM direct detection.
\end{itemize}

\subsection{The \boldmath$\eta$ case}

For the case when the $\eta$ scalar is the lightest (``$\eta$ case''), from figure~\ref{Fig:Omega} it looks as if there could be solutions for the range of $m_\eta \in [2,\,120]~\text{GeV}$. However, in the low-mass range of this interval, the $h$ invisible branching ratio, together with the relic DM density constraint becomes incompatible with the experimental data. In light of this fact, in the remainder of the discussion presented in this paragraph we shall focus on $\eta$ masses up to 120~GeV since we already know that criterion ({\it a}\,) excludes higher masses. Both constraints, i.e., ({\it a}\,) and ({\it b}\,), alongside with Cut~1 and Cut~2, are satisfied within the region $m_\eta \in [40,\,120]~\text{GeV}$. The final checks are then the di-photon ({\it c}\,) and the direct detection constraints ({\it d}\,). These two constraints are very severe and eliminate a large region of the parameter space. When imposed separately, the strongest constraint comes from the direct detection criteria, which are satisfied in the mass region $m_\eta \in [43,\,120]$~GeV and at values below $m_\eta \lesssim 10~\text{GeV}$. 

\begin{figure}[h]
\begin{center}
\hspace*{-10pt}\includegraphics[scale=0.4]{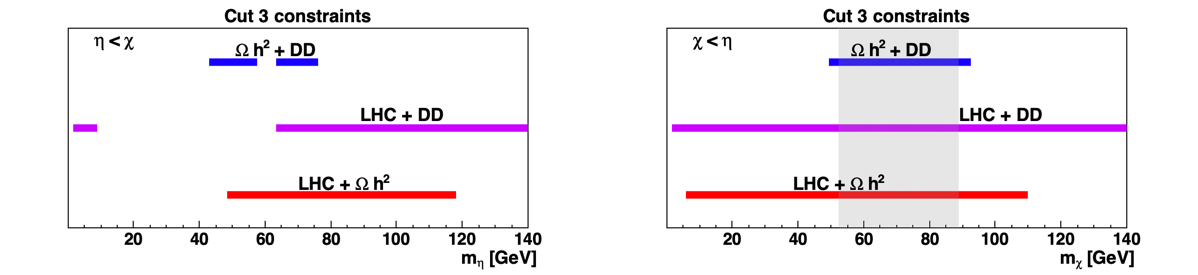}																
\end{center}
\vspace*{-4mm}
\caption{Allowed mass regions of the DM candidate involving different Cut~3 constraints. Blue: relic density satisfied together with direct detection. Purple: LHC Higgs constraints along with direct detection constraints. Red: relic density and LHC Higgs constraints. Grey: all of the Cut~3 constraints are satisfied. Note that additional input parameters are not shown. }
\label{Fig:Cut3_constraints}
\end{figure}

We list different Cut~3 paired constraints in figure~\ref{Fig:Cut3_constraints} after imposing cuts~1 and 2. We work with 8 input parameters, 6 masses and 2 angles. After imposing Cut~3, there will be different domains allowed by each of the three checks, $\Omega h^2$, DD, or LHC, imposed separately. The intersection of all these domains would correspond to Cut~3 being satisfied. Figure~\ref{Fig:Cut3_constraints} shows the allowed mass regions of the DM candidate after imposing two of the different checks at a time. Overlapping lines do not guarantee that there are regions of parameters satisfying all of the constraints simultaneously since there are seven more parameters to consider. There is no region for either ($\Omega h^2$+DD) or (LHC+DD) satisfied for $m_\eta \approx m_h/2$. In fact, for the $\eta$ case we found no parameter point satisfying all of the Cut~3 constraints simultaneously. 

\subsection{The \boldmath$\chi$ case}

For the case when the $\chi$ scalar is the lightest (``$\chi$ case''), the $\Omega h^2$ distribution is slightly shifted towards lower relic density values, as shown in figure~\ref{Fig:Omega}. As a result, the region compatible with the relic density is $m_\chi \in [2,\,105]~\text{GeV}$. For the $\eta$ case with masses below 40~GeV, when the ({\it b}\,) constraint is imposed, all $\Omega h^2$ are above 0.22. This does not apply to the $\chi$ case since in this case $\Omega h^2$ can go as low as $\approx 0.07$. Nevertheless, the sub-40 GeV region is not compatible with Cut~3. However, the region $m_\chi \in [52.5,\,89]~\text{GeV}$ survives cuts~1 to 3 when applied simultaneously. The lower DM mass range is compatible with other models presented in figure~\ref{Fig:mass-ranges}, while slightly heavier DM candidates are also allowed within the R-II-1a framework. When applied simultaneously, conditions ({\it a}\,)  and ({\it d}\,)  are satisfied for a broader range of $m_\chi \in [45.5,\,92]~\text{GeV}$. Bounds from pairs of different Cut~3 checks are shown in figure~\ref{Fig:Cut3_constraints}.

It is instructive to see which points within the parameter range survive all the cuts. The mass scatter plots of Cut~3 superimposed on the Cut~2 points are given in figure~\ref{Fig:R-II-1a-masses-exp_constraints}. There are no solutions with active neutral states being degenerate. Moreover, these masses reach at most $m_A^\mathrm{max} \sim 340~\text{GeV}$ and $m_H^\mathrm{max} \sim 450~\text{GeV}$. The CP-odd state, $A$, can be as light as the observed SM-like Higgs boson, $m_A \simeq m_h$. On the other hand, such low masses for the $H$ boson are disfavoured by Cut~3. The charged bosons that survive Cut~3 are also light, with $m_{H^\pm}^\mathrm{max} \sim 460~\text{GeV}$ and $m_{h^\pm}^\mathrm{max} \sim 310~\text{GeV}$. It is interesting to note that the charged active scalar can be as light as $m_{H^\pm}^\mathrm{min} \sim 177~\text{GeV}$.
Future experimental data on the decays of the charged bosons will be useful to test the model. For the $B$ physics constraints we applied only the indirect experimental bounds on the $\bar B \to X(s)\gamma$ rate, however other channels might lead to stronger constraints on the mass of the $H^\pm$ scalar. Finally, as noted earlier we have for the DM candidate $m_\chi \in [52.5,\,89]~\text{GeV}$} and the other inert neutral state, $\eta$, can be as heavy as $m_{\eta}^\mathrm{max} \sim 310~\text{GeV}$. It is also interesting to note that either $\eta$ or $h^\pm$ can be the next-lightest member of the inert doublet.

\begin{figure}[h]
\begin{center}
\includegraphics[scale=0.35]{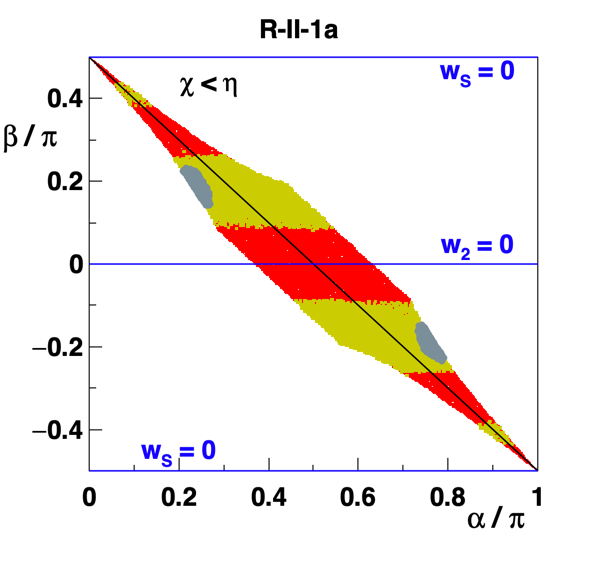}
\includegraphics[scale=0.35]{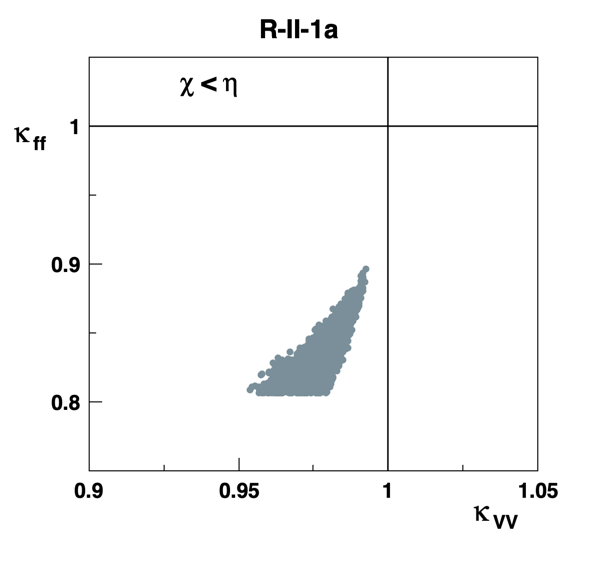}
\end{center}
\vspace*{-4mm}
\caption{Left: Constraints on $\alpha$ and $\beta$ from the gauge and Yukawa couplings. Identical colour convention and notation as in figure~\ref{Fig:SM-Like_Limit_R-II-1a}. The superimposed grey points also satisfy the Cut~3 constraints for the $\chi$ case.
Right: Corresponding constraints on relative Yukawa $\kappa_{ff}$ and gauge couplings $\kappa_{VV}$ with respect to the SM.}
\label{Fig:SM-Like_Limit_R-II-1a-Cut3}
\end{figure}

In addition to the mass parameters, angles are also used as input.
The allowed ranges in the $\alpha\,$-$\,\beta$ plane are shown in figure~\ref{Fig:SM-Like_Limit_R-II-1a-Cut3}. The solutions satisfying all cuts are asymmetrically distributed along the black diagonal line, which represents the SM-like limit. In the right panel of figure~\ref{Fig:SM-Like_Limit_R-II-1a-Cut3} we also show the gauge ($\kappa_{VV}$) and Yukawa ($\kappa_{ff}$) couplings, relative to the SM values, see eqs.~\eqref{Eq:RII1aSMLikeLimit}. Figure~\ref{Fig:SM-Like_Limit_R-II-1a-Cut3} shows that the experimental data are more stringent for $\kappa_{ff}$ than for $\kappa_{VV}$. The lack of symmetry in the distribution of the grey points with respect to the diagonal, corrresponding to the SM-like limit, in the left panel translates into a significant population of $\kappa_{ff}$-values below unity in the right panel.

\begin{figure}[htb]
\begin{center}
\includegraphics[scale=0.25]{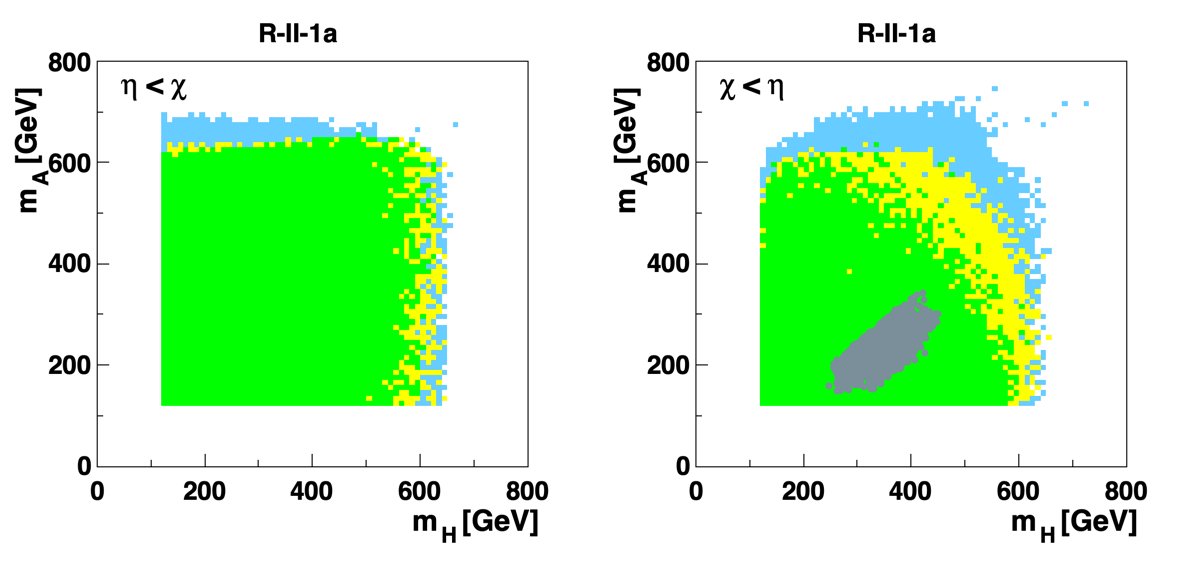}
\includegraphics[scale=0.25]{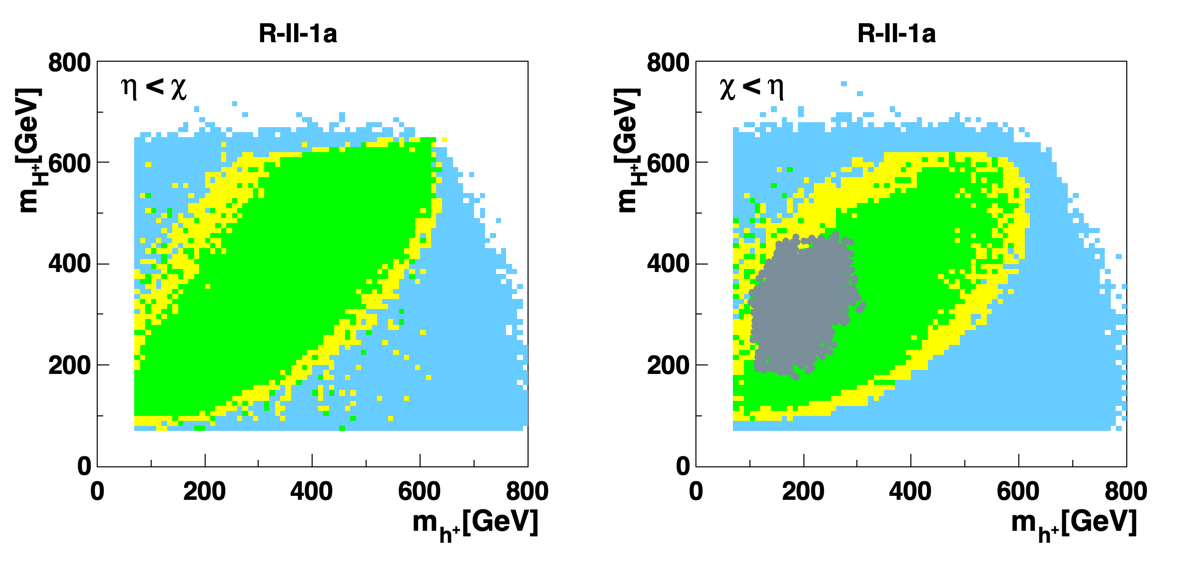}
\includegraphics[scale=0.25]{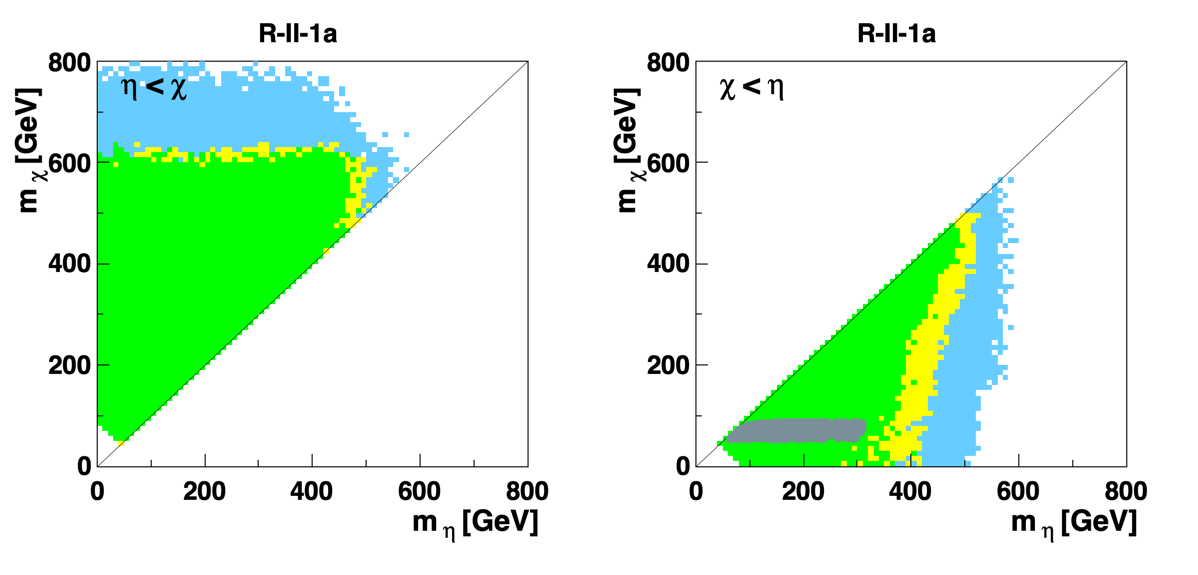}
\end{center}
\vspace*{-10mm}
\caption{Scatter plots of masses that satisfy different cuts, for both orderings of $\eta$ and $\chi$. Identical notation as in figure~\ref{Fig:R-II-1a-masses-HA_and_charged-th_constraints}. The light-blue region satisfies Cut~1 and accommodates the $16\pi$ unitarity constraint. The yellow region accommodates a 3-$\sigma$ tolerance with respect to Cut~2,  whereas in the green regions, the model is within the 2-$\sigma$ bound of these values. The grey points are compatible with all cuts and are only present in the right-hand~panels.}
\label{Fig:R-II-1a-masses-exp_constraints}
\end{figure}

\begin{figure}[h]
\begin{center}
\includegraphics[scale=0.4]{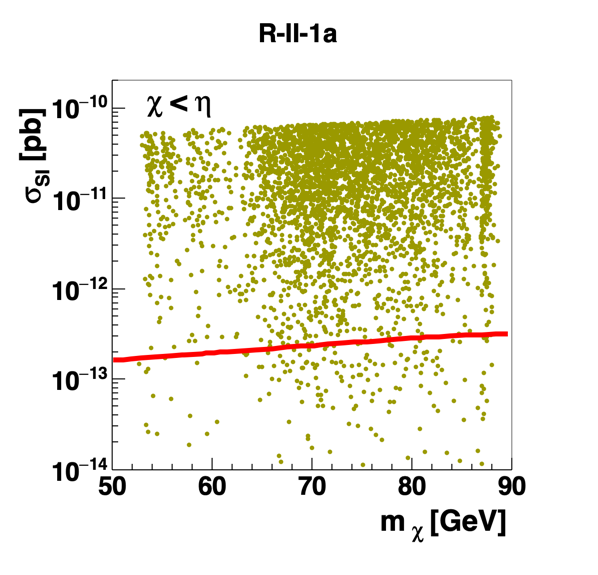}
\end{center}
\vspace*{-4mm}
\caption{The spin-independent DM-nucleon cross section compatible with XENON1T~\cite{Aprile:2018dbl} data at 90\% C.L. The points represent Cut~3 satisfied for the $\chi$ DM case. The red line corresponds to an approximate neutrino floor, which can be defined in various ways. }
\label{Fig:Xenon1T}
\end{figure}

We present direct detection constraints in figure~\ref{Fig:Xenon1T}. We note that in practically the whole mass range there are parameter points at lower cross sections. Thus, a future improvement on this direct detection constraint is not obviously going to reduce the range of masses allowed by the model.

Finally, in table~\ref{Table:benchmarks} we show some benchmarks.

{\renewcommand{\arraystretch}{1.4}
\begin{table}[htb] \footnotesize
\caption{Benchmark points and dominant decay modes. The ``$q$'' notation refers to a sum over the light quarks, $d$, $u$, $s$ and $c$, ``$l$'' refers to all leptons, and ``$\nu$'' to all neutrinos.}
\label{Table:benchmarks}
\begin{center}
\resizebox{14cm}{!}{
\begin{tabular}{|c|c|c|c|c|c|c|c|c|c|}
\hline\hline
Parameter & BP 1 & BP 2 & BP 3 & BP 4 & BP 5 & BP6 & BP7 & BP8 & BP9 \\
\hline
\hline
DM ($\chi$) mass [GeV] & 52.6 & 56.1 & 59.6 & 63.02 & 65.7 & 70.3 & 75.0 & 82.2 & 88.6 \\ \hline
$\eta$ mass [GeV] & 62.7 & 203.8 & 270.4 & 169.4 & 150.5 & 157.7 & 202.8 & 127.8 & 210.7 \\ \hline
$h^+$ mass [GeV] & 115.4 & 167.4 & 273.6 & 188.6 & 214.1 & 170.5 & 232.0 & 151.8 & 243.0 \\ \hline
$H^+$ mass [GeV] & 192.6 & 369.5 & 367.4 & 246.6 & 265.5 & 405.8 & 319.8 & 410.6 & 311.9 \\ \hline
$H$ mass [GeV] & 263.9 & 349.3 & 352.9 & 276.3 & 298.2 & 402.0 & 368.5 & 405.2 & 317.6 \\ \hline
$A$ mass [GeV] & 179.2 & 208.0 & 190.7 & 173.9 & 205.2 & 255.3 & 251.3 & 330.0 & 247.0 \\ \hline
$\beta/\pi$ & 0.162 & -0.204 & -0.201 & -0.165 & 0.163 & 0.220 & 0.203 & -0.218 & 0.183 \\ \hline
$\alpha/\pi$ & 0.252 & 0.763 & 0.765 & 0.752 & 0.254 & 0.225 & 0.239 & 0.769 & 0.238 \\ \hline
$\sigma_\text{SI}\, [10^{-11}~\text{pb}]$ & 0.029 & 1.456 & 4.928 & 0.176 & 5.326 & 1.341 & 2.711 & 8.553 & 4.491 \\ \hline\hline

$\eta\to\chi q\bar q$ [\%] & 63.27 &  &  &  & 54.38 & 54.35 &  & 53.95 &  \\ \hline
$\eta\to\chi b\bar b$ [\%] & 0.48 &  &  &  & 14.80 & 14.85 &  & 13.90 &  \\ \hline
$\eta\to\chi \nu \bar\nu$ [\%] & 24.62 &  &  &  & 20.48 & 20.46 &  & 20.72 &  \\ \hline
$\eta\to\chi l \bar l$ [\%]  & 11.61 &  &  &  & 10.33 & 10.33 &  & 11.42 &  \\ \hline
$\eta\to\chi Z $ [\%]  &  & 99.98 & 53.09 & 100 &  &  & 100 &  & 100 \\ \hline
$\eta\to\chi A $ [\%] &  &  & 46.91 &  &  &  &  &  &  \\ \hline \hline

$h^+\to\chi W^+$ [\%]  &  & 100 & 100 & 99.98 & 99.89 & 99.99 & 99.99 &  & 99.99 \\ \hline
$h^+\to\eta q\bar q$ [\%]  & 20.18 &  &  &  &  &  &  & 0.30 &  \\ \hline
$h^+\to\eta \nu \bar l$ [\%] & 9.88 &  &  &  &  &  &  & 0.16 &  \\ \hline
$h^+\to\chi q\bar q$ [\%]  & 46.94  &  &  &  &  &  &  & 66.82 &  \\ \hline
$h^+\to\chi \nu \bar l$ [\%] & 22.99 &  &  &  &  &  &  & 32.71 &  \\\hline\hline

$H^+\to t\bar b$ [\%]  & 9.07 & 43.69 & 58.23 & 95.09 & 95.78 & 30.95 & 96.25 & 31.54 & 93.59 \\ \hline
$H^+\to AW^+$ [\%]  &  & 20.56 & 35.74 & 0.29 & 0.06 & 8.66 & 0.05 & 0.05 & 0.05 \\ \hline
$H^+\to hW^+$ [\%]  &  & 1.94 &2.67  &  4.46 & 4.00 & 1.23 & 2.86 & 1.15 & 6.20 \\ \hline
$H^+\to h^+ \eta$ [\%]  & 85.9 &  &  &  &  & 43.74 &  & 61.68 &  \\ \hline
$H^+\to h^+ \chi$ [\%]  & 5.00  & 33.74 & 3.26 &  &  & 15.36  & 0.68 & 5.53 &  \\ \hline \hline

$H\to\chi \chi$ [\%]  &  0.15 & 0.03 & 0.07 & 0.87 & 15.03 &  & 11.34 & 7.63 & 63.75 \\ \hline
$H\to\eta\eta$ [\%]  & 89.9 &  &  &  &  & 24.89 &  & 25.31 &  \\ \hline
$H\to hh$ [\%]  & 3.07 & 2.64 & 9.40 & 34.59 & 33.53 & 1.33 & 13.43 & 0.88 & 14.72 \\ \hline
$H\to AZ$ [\%]  & 0.09 & 13.55 & 70.93 & 13.91 & 2.87 & 7.61 & 22.78 &  & 0.07 \\ \hline
$H\to W^+W^-$ [\%]  & 4.06 & 3.13 & 10.40 & 34.98 & 33.35 & 1.89 & 16.32 & 1.26 & 14.70 \\ \hline
$H\to Z Z$ [\%]  & 1.75 & 1.43 & 4.77 & 15.29 & 14.82 & 0.88 & 7.53 & 0.59 & 6.62 \\ \hline
$H\to h^+ h^-$ [\%]  & 0.80 & 78.59 &  &  &  & 52.94 &  & 56.33 &  \\ \hline
$H\to q \bar q$ [\%]  &  & 0.62 & 4.40 & 0.32 & 0.34 & 10.43 & 28.52 & 8.00 & 0.12 \\ \hline \hline

$A\to \eta \chi$ [\%]  & 99.97 &  &  &  &  & 99.32 &  & 99.01 &  \\ \hline
$A\to b\bar b$ [\%]  & 0.02 & 79.78 & 84.15 & 84.63 & 75.28 & 0.07 & 8.84 & 0.02 & 4.76 \\ \hline
$A\to q\bar q$ [\%]  &  & 3.56 & 3.75 & 3.77 & 3.36 &  & 0.39 &  & 0.21 \\ \hline
$A\to \tau^+\tau^-$ [\%]  &  & 9.85 & 10.19 & 10.00 & 9.24 &  & 1.13 &  & 0.61 \\ \hline
$A\to h Z$ [\%]  &  & 6.81 & 1.87 & 1.55 & 12.08 & 0.60 & 89.63 & 0.96 & 94.42 \\ \hline\hline
\end{tabular}
}
\end{center}
\end{table}

\clearpage

\section{Concluding remarks}
\label{sect:conclude}

It is possible to incorporate a DM candidate within the $S_3$-symmetric scalar model. The DM candidate requires one of the Higgs doublets to be inert. As $S_3$ symmetry is assumed and a DM candidate is sought, this requirement imposes constraints on the structure of the Yukawa Lagrangian. As a matter of fact, we found no possible combinations of the exact $S_3$-symmetric scalar potential and a non-trivial Yukawa Lagrangian, which could accommodate a DM candidate. Hence we require fermions to transform trivially under $S_3$. When soft-symmetry breaking terms are present, it is possible to construct the Yukawa Lagrangian with a non-trivial $S_3$ structure. Due to such behaviour an \textit{ad-hoc} $S_3$ is not appealing in the context of being simultaneously applied to the scalar sector and generating a non-trivial Yukawa sector, while trying to explain DM.

In this work we focused on a specific $S_3$-symmetric scalar model R-II-1a. As was shown in the paper, this model is compatible with several constraints, both theoretical and experimental. In the IDM and the literature on 3HDMs there is a viable DM high-mass region present. This is not true in our case. The main difference is that the inert-active scalar couplings in the R-II-1a are constrained by the underlying $S_3$ symmetry and hence the portal couplings are harder to adjust.

We analysed the R-II-1a model numerically. Within the model there are two possible DM candidates present, $\eta$ and $\chi$. After performing the analysis we found no solutions satisfying all of the constraints with $\eta$ being the lightest. However, for the $\chi$ case there is a range compatible with all constraints, $m_\chi \in [52.5,\,89]~\text{GeV}$. Constraints in this mass range are compatible with data at the 3-$\sigma$ level. As compared with the IDM, the R-II-1a model allows solutions in the intermediate-mass range up to somewhat higher values, but can not satisfy all constraints in the high-mass region.

The model differs from the IDM in having two non-inert doublets. The corresponding scalars must be rather light, as shown in figure~\ref{Fig:R-II-1a-masses-exp_constraints}. If these were to be produced at the LHC, they could decay to a pair of scalars from the inert doublet, as well as to final states familiar from the 2HDM. Furthermore, one could imagine direct production of two scalars of the inert doublet, for example via $WW$ or $WZ$ fusion, at rates controlled by the gauge couplings. For all of these cases, the charged $h^\pm$ and neutral $\eta$ would decay via gauge boson emission to the DM, $h^\pm\to W^\pm\chi$ and $\eta\to Z\chi$, leading to mono-vector events~\cite{ATLAS:2018nda,CMS:2021snz}.

The present model  looks very promising. There are some similarities between this model and the 2HDM. The present framework preserves CP both at the Lagrangian level and by the vacuum. The presence of both $g\left(\eta h^\pm H^\mp\right)$ and $g\left(\chi h^\pm H^\mp\right)$ couplings suggests there might be mixing at the one-loop level, but the two diagrams associated with the different charge assignments cancel.
Since there is no CP violation in the scalar sector, the Yukawa couplings must be complex and CP is only violated via the CKM matrix. The  fermions only couple to one of the active scalars, the one that is a singlet under $S_3$ since the fermions also transform trivially under $S_3$. In the $S_3$-symmetric 3HDM there are also regions of parameter space leading to vacua that violate CP spontaneously, as discussed in Refs.~\cite{Emmanuel-Costa:2016vej,Kuncinas:2020wrn}, showing that the $S_3$-symmetric 3HDM has a very rich structure.

\section*{Acknowledgements}

It is a pleasure to thank Igor Ivanov, Mikolaj Misiak and Alexander Pukhov for very useful discussions. We thank the referee for pointing out a mistake in the earlier version.
PO~is supported in part by the Research Council of Norway.
The work of AK and MNR was partially supported by Funda\c c\~ ao 
para  a  Ci\^ encia e a Tecnologia  (FCT, Portugal)  through  the  projects  
CFTP-FCT Unit UIDB/00777/2020 and UIDP/00777/2020, CERN/FIS-PAR/0008/2019 and 
PTDC/FIS-PAR/29436/2017  which are  partially  funded  through
POCTI  (FEDER),  COMPETE,  QREN  and  EU. Furthermore, the work of AK has been supported by the FCT PhD fellowship with reference UI/BD/150735/2020.
WK would like to thank the Associateship Scheme at the Abdus Salam International Centre for Theoretical Physics (ICTP) for their support.
MNR and PO benefited from discussions that took place at the University of Warsaw during visits supported by the HARMONIA project of the National Science Centre, Poland, under contract
UMO-2015/18/M/ST2/00518 (2016-2019),
 both also thank the University of Bergen  and
CFTP/ IST/University of Lisbon, where collaboration visits took place. 

\appendix
\section{Scalar-scalar couplings}
\label{App:Scalar_Couplings_RII1a}
\setcounter{equation}{0}

For simplicity, the scalar-scalar couplings are presented with the symmetry factor, but without the overall coefficient ``$-i$". We denote the ``correct" couplings as $g_{\dots} = -i g\left({\dots}\right)$. We shall abbreviate $\mathrm{c}_\theta\equiv\cos\theta$, and $\mathrm{s}_\theta\equiv\sin\theta$, and $\mathrm{t}_\theta\equiv\tan\theta$ for any argument $\theta$.

The trilinear scalar-scalar couplings involving the same species are:
\begin{subequations}
\begin{align}
\begin{split}
g\left( h h h \right) & = 3 v \bigg[\mathrm{c}_{\alpha }^3 \left(2 \left(\lambda _1+\lambda _3\right) \mathrm{s}_{\beta }-\lambda _4 \mathrm{c}_{\beta }\right)+ \left(\lambda _a \mathrm{c}_{\beta }-3 \lambda _4 \mathrm{s}_{\beta }\right)\mathrm{c}_{\alpha }^2 \mathrm{s}_{\alpha }\\&\hspace{30pt}+\lambda _a \mathrm{c}_{\alpha }\mathrm{s}_{\alpha }^2 \mathrm{s}_{\beta } +2 \lambda _8 \mathrm{s}_{\alpha }^3 \mathrm{c}_{\beta }\bigg]\label{Eq.R_II_1a_hhh},
\end{split}\\
\begin{split}
g\left( H H H \right) & =-3 v \bigg[\mathrm{s}_{\alpha }^3 \left(2 \left(\lambda _1+\lambda _3\right) \mathrm{s}_{\beta }-\lambda _4 \mathrm{c}_{\beta }\right)+ \left(3 \lambda _4 \mathrm{s}_{\beta }-\lambda _a \mathrm{c}_{\beta }\right)\mathrm{c}_{\alpha } \mathrm{s}_{\alpha }^2 \\&\hspace{40pt}+\lambda _a \mathrm{c}_{\alpha }^2\mathrm{s}_{\alpha } \mathrm{s}_{\beta } -2 \lambda _8 \mathrm{c}_{\alpha }^3 \mathrm{c}_{\beta }\bigg].
\end{split}
\end{align} 
\end{subequations}
The trilinear couplings involving the neutral fields are:
\begin{subequations}
\begin{align}
g\left( \eta  \eta  h \right) & = v \left[\mathrm{s}_{\alpha } \left(3 \lambda _4 \mathrm{s}_{\beta } + \lambda _a \mathrm{c}_{\beta }\right)+\mathrm{c}_{\alpha } \left(2 \left(\lambda _1+\lambda _3\right) \mathrm{s}_{\beta }+3 \lambda _4 \mathrm{c}_{\beta }\right)\right]\label{Eq.R_II_1a_hetaeta},\\
g\left( \eta  \eta  H \right) & = v \left[ \mathrm{c}_{\alpha } \left(3 \lambda _4 \mathrm{s}_{\beta } + \lambda _a \mathrm{c}_{\beta }\right)- \mathrm{s}_{\alpha } \left(2 \left(\lambda _1+\lambda _3\right) \mathrm{s}_{\beta }+3 \lambda _4 \mathrm{c}_{\beta }\right) \right],\\
g\left( \chi  \chi  h \right) & = v \left[\mathrm{s}_{\alpha } \left(\lambda _4 \mathrm{s}_{\beta } + \lambda _b \mathrm{c}_{\beta }\right)+\mathrm{c}_{\alpha } \left(2 \left(\lambda _1-2 \lambda _2-\lambda _3\right) \mathrm{s}_{\beta }+\lambda _4 \mathrm{c}_{\beta }\right)\right]\label{Eq.R_II_1a_hchichi},\\
g\left( \chi  \chi  H \right) & = v \left[\mathrm{c}_{\alpha } \left(\lambda _4 \mathrm{s}_{\beta } + \lambda _b \mathrm{c}_{\beta }\right)-\mathrm{s}_{\alpha } \left(2 \left(\lambda _1-2 \lambda _2-\lambda _3\right) \mathrm{s}_{\beta }+\lambda _4 \mathrm{c}_{\beta }\right)\right],\\
g\left( \eta  \chi  A \right) & = -v \left[ \lambda _4 \mathrm{c}_{2 \beta }+\left(\lambda _2+\lambda _3-\lambda _7\right) \mathrm{s}_{2 \beta } \right],\\
\begin{split}
g\left( h h H \right) & = -v \bigg[\mathrm{c}_{\alpha }^3 \left(3 \lambda _4 \mathrm{s}_{\beta }-\lambda _a \mathrm{c}_{\beta }\right)+\mathrm{c}_{\alpha }^2 \mathrm{s}_{\alpha } \left(2\left(3 \lambda _1 + 3\lambda _3-\lambda _a\right) \mathrm{s}_{\beta }-3 \lambda _4 \mathrm{c}_{\beta }\right)\\& \hspace{35pt}+\lambda _a \mathrm{s}_{\alpha }^3 \mathrm{s}_{\beta }  - 2 \mathrm{c}_{\alpha } \mathrm{s}_{\alpha }^2 \left(3 \lambda _4 \mathrm{s}_{\beta }+\left(3 \lambda _8-\lambda _a\right) \mathrm{c}_{\beta }\right)\bigg],
\end{split}\\
\begin{split}
g\left( h H H \right) & = v \bigg[-\mathrm{s}_{\alpha }^3 \left(3 \lambda _4 \mathrm{s}_{\beta }-\lambda _a \mathrm{c}_{\beta }\right)+\mathrm{c}_{\alpha } \mathrm{s}_{\alpha }^2 \left(2\left(3 \lambda _1 + 3\lambda _3-\lambda _a\right) \mathrm{s}_{\beta }-3 \lambda _4 \mathrm{c}_{\beta }\right)\\& \hspace{30pt}+\lambda _a \mathrm{c}_{\alpha }^3 \mathrm{s}_{\beta }  + 2 \mathrm{c}_{\alpha }^2 \mathrm{s}_{\alpha } \left(3 \lambda _4 \mathrm{s}_{\beta }+\left(3 \lambda _8-\lambda _a\right) \mathrm{c}_{\beta }\right)\bigg],
\end{split}\\
\begin{split}
g\left( A A h \right) & = v \bigg[ \left( \lambda_4 \left( 2 \mathrm{c}_{\beta } \mathrm{s}_{\beta }^2 - \mathrm{c}_{\beta }^3 \right) + \lambda _b \mathrm{s}_{\beta }^3 +2\left(\lambda _1+\lambda _3-2 \lambda _7\right) \mathrm{c}_{\beta }^2 \mathrm{s}_{\beta }\right)\mathrm{c}_{\alpha }\\&\hspace{30pt}+\left(
-\frac{1}{2} \lambda _4 \mathrm{s}_{2 \beta } -2 \left(2 \lambda _7 - \lambda _8\right) \mathrm{s}_{\beta }^2 + \lambda _b \mathrm{c}_{\beta }^2 \right)\mathrm{s}_{\alpha } \mathrm{c}_{\beta }  \bigg],
\end{split}\\
\begin{split}
g\left( A A H \right) & = v \bigg[ -\left( \lambda_4 \left( 2 \mathrm{c}_{\beta } \mathrm{s}_{\beta }^2 - \mathrm{c}_{\beta }^3 \right) + \lambda _b \mathrm{s}_{\beta }^3 +2\left(\lambda _1+\lambda _3-2 \lambda _7\right) \mathrm{c}_{\beta }^2 \mathrm{s}_{\beta }\right)\mathrm{s}_{\alpha }\\&\hspace{30pt}+\left(
-\frac{1}{2} \lambda _4 \mathrm{s}_{2 \beta } -2 \left(2 \lambda _7 - \lambda _8\right) \mathrm{s}_{\beta }^2 + \lambda _b \mathrm{c}_{\beta }^2 \right)\mathrm{c}_{\alpha } \mathrm{c}_{\beta }  \bigg].
\end{split}
\end{align} 
\end{subequations}
The trilinear couplings involving the charged fields are:
\begin{subequations}
\begin{align}
g\left( \eta  h^\pm H^\mp \right) & = -\frac{1}{4} v \left[4 \lambda _4 \mathrm{c}_{2 \beta }+\left(4 \lambda _3-\lambda _6-2 \lambda _7\right) \mathrm{s}_{2 \beta }\right],\\
g\left( \chi  h^\pm H^\mp \right) & = \mp\frac{1}{4} i v \left(4 \lambda _2+\lambda _6-2 \lambda _7\right) \mathrm{s}_{2 \beta },\\
\begin{split}
g\left( h H^\pm H^\mp \right) & = -v \bigg[\left( \lambda_4 \left( \mathrm{c}_{\beta }^3 -2 \mathrm{c}_{\beta } \mathrm{s}_{\beta }^2\right)-\lambda _5 \mathrm{s}_{\beta }^3-\left(2 \lambda_1+2 \lambda_3-\lambda _6-2 \lambda _7\right) \mathrm{c}_{\beta }^2 \mathrm{s}_{\beta }\right)\mathrm{c}_{\alpha} \\&\hspace{40pt}+\left( \lambda _4 \mathrm{c}_{\beta }^2\mathrm{s}_{\beta }-\lambda _5 \mathrm{c}_{\beta }^3 + \lambda_7 \mathrm{s}_{\beta } \mathrm{s}_{2\beta } + \left( \lambda_6 - 2 \lambda_8 \right) \mathrm{c}_{\beta } \mathrm{s}_{\beta }^2 \right)\mathrm{s}_{\alpha } \bigg],
\end{split}\\
\begin{split}
g\left( H H^\pm H^\mp \right) & = v \bigg[\left( \lambda_4 \left( \mathrm{c}_{\beta }^3 -2 \mathrm{c}_{\beta } \mathrm{s}_{\beta }^2\right)-\lambda _5 \mathrm{s}_{\beta }^3-\left(2 \lambda_1+2 \lambda_3-\lambda _6-2 \lambda _7\right) \mathrm{c}_{\beta }^2 \mathrm{s}_{\beta }\right)\mathrm{s}_{\alpha} \\&\hspace{30pt}-\left( \lambda _4 \mathrm{c}_{\beta }\mathrm{s}_{\beta }-\lambda _5 \mathrm{c}_{\beta }^2 + \left( \lambda_6 + 2 \lambda_7 - 2 \lambda_8 \right) \mathrm{s}_{\beta }^2 \right)\mathrm{c}_{\alpha }\mathrm{c}_{\beta } \bigg],
\end{split}\\
g\left( h h^\pm h^\mp \right) & = v \left[\mathrm{c}_{\alpha } \left(2 \left(\lambda _1-\lambda _3\right) \mathrm{s}_{\beta }+\lambda _4 \mathrm{c}_{\beta }\right)+\mathrm{s}_{\alpha } \left(\lambda _4 \mathrm{s}_{\beta }+\lambda _5 \mathrm{c}_{\beta }\right)\right],\\
g\left( H h^\pm h^\mp \right) & = v \left[ - \mathrm{s}_{\alpha } \left(2 \left(\lambda _1-\lambda _3\right) \mathrm{s}_{\beta }+\lambda _4 \mathrm{c}_{\beta }\right) + \mathrm{c}_{\alpha } \left(\lambda _4 \mathrm{s}_{\beta }+\lambda _5 \mathrm{c}_{\beta }\right)\right].
\end{align} 
\end{subequations}

The quartic couplings involving the same species are:
\begin{subequations}
\begin{align}
g\left( \eta  \eta  \eta  \eta  \right) & = g\left( \chi  \chi  \chi  \chi  \right) = 6 \left(\lambda _1+\lambda _3\right)\label{Eq.R_II_1a_QCLim_l1l3},\\
g\left( h h h h \right) & = 6 \left[\left(\lambda _1+\lambda _3\right) \mathrm{c}_{\alpha }^4-2 \lambda _4 \mathrm{c}_{\alpha }^3 \mathrm{s}_{\alpha }+\lambda _8 \mathrm{s}_{\alpha }^4 + \frac{1}{4}\lambda _a \mathrm{s}_{2\alpha }^2 \right]\label{Eq.R_II_1a_hhhh},\\
g\left( H H H H \right) & = 6 \left[\left(\lambda _1+\lambda _3\right) \mathrm{s}_{\alpha }^4 + 2 \lambda _4 \mathrm{c}_{\alpha } \mathrm{s}_{\alpha }^3 + \lambda _8 \mathrm{c}_{\alpha }^4 + \frac{1}{4}\lambda _a \mathrm{s}_{2\alpha }^2\right],\\
g\left( A A A A \right) &= 6 \left[ \left(\lambda _1+\lambda _3\right) \mathrm{c}_{\beta}^4 + 2 \lambda _4 \mathrm{c}_{\beta}^3 \mathrm{s}_{\beta} + \lambda _8 \mathrm{s}_{\beta }^4 + \frac{1}{4}\lambda _a \mathrm{s}_{2\beta }^2
\right].
\end{align} 
\end{subequations}
The quartic couplings involving only the neutral fields are:
\begin{subequations}
\begin{align}
g\left( \eta  \eta  \chi  \chi  \right) & = 2 \left(\lambda _1+\lambda _3\right),\\
g\left( \eta  \eta  A A \right) & = 2 \left(\lambda _1-2 \lambda _2-\lambda _3\right) \mathrm{c}_{\beta }^2-\lambda _4 \mathrm{s}_{2 \beta }+\lambda _b \mathrm{s}_{\beta }^2,\\
g\left( \chi  \chi  A A \right) & = 2 \left(\lambda _1+\lambda _3\right) \mathrm{c}_{\beta }^2-3 \lambda _4 \mathrm{s}_{2 \beta }+\lambda _a \mathrm{s}_{\beta }^2,\\
g\left( \eta  \eta  h h \right) & = 2 \left(\lambda _1+\lambda _3\right) \mathrm{c}_{\alpha }^2+3 \lambda _4 \mathrm{s}_{2\alpha }+\lambda _a \mathrm{s}_{\alpha }^2,\\
g\left( \eta  \eta  h H \right) & = -\frac{1}{2}\left(2 \lambda _1+2 \lambda _3 - \lambda _a \right) \mathrm{s}_{2\alpha }+3 \lambda _4 \mathrm{c}_{2 \alpha } ,\\
g\left( \eta  \eta  H H \right) & = 2 \left(\lambda _1+\lambda _3\right) \mathrm{s}_{\alpha }^2-3 \lambda _4 \mathrm{s}_{2\alpha }+\lambda _a \mathrm{c}_{\alpha }^2,\\
g\left( \chi  \chi  h h \right) & = 2 \left(\lambda _1-2 \lambda _2-\lambda _3\right) \mathrm{c}_{\alpha }^2+ \lambda _4 \mathrm{s}_{2\alpha }+\lambda _b \mathrm{s}_{\alpha }^2,\\
g\left( \chi  \chi  h H \right) & = \lambda _4 \mathrm{c}_{2 \alpha }-\frac{1}{2}\left(2 \lambda _1-4 \lambda _2-2 \lambda _3-\lambda _b\right) \mathrm{s}_{2\alpha },\\
g\left( \chi  \chi  H H \right) & = 2 \left(\lambda _1-2 \lambda _2-\lambda _3\right) \mathrm{s}_{\alpha }^2-\lambda _4 \mathrm{s}_{2 \alpha }+\lambda _b \mathrm{c}_{\alpha }^2,\\
g\left( \eta  \chi  h A \right) & = -\mathrm{c}_{\alpha } \left[2 \left(\lambda _2+\lambda _3\right) \mathrm{c}_{\beta }-\lambda _4 \mathrm{s}_{\beta }\right]-\left(\lambda _4 \mathrm{c}_{\beta }-2 \lambda _7 \mathrm{s}_{\beta }\right)\mathrm{s}_{\alpha },\\
g\left( \eta  \chi  H A \right) & = \mathrm{s}_{\alpha } \left[2 \left(\lambda _2+\lambda _3\right) \mathrm{c}_{\beta }-\lambda _4 \mathrm{s}_{\beta }\right]- \left(\lambda _4 \mathrm{c}_{\beta }-2 \lambda _7 \mathrm{s}_{\beta }\right)\mathrm{c}_{\alpha },\\
g\left( h h h H \right) & = -3 \mathrm{c}_{\alpha } \left[\lambda _4 \mathrm{c}_{3 \alpha }+ \left(\lambda _1+\lambda _3-\lambda _8+ \left(\lambda _1+\lambda _3-\lambda_a+\lambda _8\right) \mathrm{c}_{2 \alpha }\right)\mathrm{s}_{\alpha }\right],\\
g\left( h h H H \right) & = \frac{1}{4} \left[3 \lambda _1+3 \lambda _3+6 \lambda _4 \mathrm{s}_{4 \alpha }+ \lambda _a+3 \lambda _8 -3 \left(\lambda _1+\lambda _3-\lambda_a+\lambda _8\right) \mathrm{c}_{4 \alpha }\right] ,\\
g\left( h H H H \right) & = -\frac{3}{2} \mathrm{s}_{\alpha } \left[2 \lambda _4 \mathrm{s}_{3 \alpha }+\left(\lambda _1+\lambda _3+\lambda _a-3 \lambda _8\right) \mathrm{c}_{\alpha }-\left(\lambda _1+\lambda _3-\lambda_a+\lambda _8\right) \mathrm{c}_{3 \alpha }\right] ,\\
\begin{split}
g\left( A A h h \right) & = - \mathrm{s}_{2\alpha} \mathrm{c}_\beta \left(\lambda _4 \mathrm{c}_{\beta }+4 \lambda _7 \mathrm{s}_{\beta }\right) + \mathrm{s}_\alpha^2 \left(\lambda_b \mathrm{c}_{\beta }^2+2 \lambda _8 \mathrm{s}_{\beta }^2\right)\\&\hspace{13pt} + \mathrm{c}_{\alpha }^2\left( 2 \left(\lambda _1+\lambda _3\right) \mathrm{c}_{\beta }^2 +\lambda _4 \mathrm{s}_{2 \beta } + \lambda _b \mathrm{s}_{\beta }^2\right),
\end{split}\\
\begin{split}
g\left( A A h H \right) & = -\frac{1}{2}\mathrm{s}_{2\alpha } \left[\lambda _1+\lambda _3+\lambda _4 \mathrm{s}_{2 \beta }-\lambda _8+\left(\lambda _1+\lambda _3-\lambda_b+\lambda _8\right) \mathrm{c}_{2 \beta }\right]\\&\hspace{13pt}-\mathrm{c}_{2\alpha } \mathrm{c}_{\beta } \left(\lambda _4 \mathrm{c}_{\beta } + 4 \lambda _7 \mathrm{s}_{\beta }\right),
\end{split}\\
\begin{split}
g\left( A A H H \right) & = \mathrm{s}_{2\alpha} \mathrm{c}_\beta \left(\lambda _4 \mathrm{c}_{\beta }+4 \lambda _7 \mathrm{s}_{\beta }\right) + \mathrm{c}_\alpha^2 \left(\lambda_b \mathrm{c}_{\beta }^2+2 \lambda _8 \mathrm{s}_{\beta }^2\right)\\&\hspace{13pt} + \mathrm{s}_{\alpha }^2\left( 2 \left(\lambda _1+\lambda _3\right) \mathrm{c}_{\beta }^2 +\lambda _4 \mathrm{s}_{2 \beta } + \lambda _b \mathrm{s}_{\beta }^2\right),
\end{split}
\end{align} 
\end{subequations}
The quartic couplings involving both neutral and charged fields are:
\begin{subequations}
\begin{align}
g\left( \eta  \eta  h^\pm h^\mp \right) & =  g\left( \chi  \chi  h^\pm h^\mp \right) = 2 \left(\lambda_1+\lambda_3\right),\\
g\left( \eta  \eta  H^\pm H^\mp \right) & = g\left( \chi  \chi  H^\mp H^\pm \right) =2 \left(\lambda_1-\lambda_3\right) \mathrm{c}_{\beta }^2-\lambda _4 \mathrm{s}_{2 \beta }+ \lambda _5 \mathrm{s}_{\beta }^2 ,\\
g\left( \eta  h h^\pm H^\mp \right) & = - \left(2 \lambda _3 \mathrm{c}_{\beta }-\lambda _4 \mathrm{s}_{\beta }\right)\mathrm{c}_{\alpha }-\lambda _4\mathrm{s}_{\alpha } \mathrm{c}_{\beta }+\frac{1}{2} \left(\lambda _6+2 \lambda _7\right) \mathrm{s}_{\alpha }\mathrm{s}_{\beta } ,\\
g\left( \eta  H h^\pm H^\mp \right) & = \left(2 \lambda _3 \mathrm{c}_{\beta }-\lambda _4 \mathrm{s}_{\beta }\right)\mathrm{s}_{\alpha }-\lambda _4 \mathrm{c}_{\alpha } \mathrm{c}_{\beta }+\frac{1}{2} \left(\lambda _6+2 \lambda _7\right) \mathrm{c}_{\alpha } \mathrm{s}_{\beta },\\
g\left( \chi  h h^\pm H^\mp \right) & = \mp i \left[2 \lambda _2 \mathrm{c}_{\alpha } \mathrm{c}_{\beta }+\frac{1}{2} \left(\lambda _6-2 \lambda _7\right) \mathrm{s}_{\alpha }\mathrm{s}_{\beta } \right],\\
g\left( \chi  H h^\pm H^\mp \right) & = \mp i \left[-2 \lambda _2 \mathrm{s}_{\alpha } \mathrm{c}_{\beta }+\frac{1}{2} \left(\lambda _6-2 \lambda _7\right) \mathrm{c}_{\alpha } \mathrm{s}_{\beta }\right],\\
g\left( h h h^\pm h^\mp \right) & =2 \left(\lambda _1-\lambda _3\right) \mathrm{c}_{\alpha }^2+\lambda _4 \mathrm{s}_{2 \alpha }+\lambda _5 \mathrm{s}_{\alpha }^2,\\
g\left( h H h^\pm h^\mp \right) & =-\frac{1}{2}\left(2\lambda _1-2\lambda _3-\lambda _5\right) \mathrm{s}_{2\alpha }+\lambda _4 \mathrm{c}_{2 \alpha },\\
g\left( H H h^\pm h^\mp \right) & =2 \left(\lambda _1-\lambda _3\right) \mathrm{s}_{\alpha }^2-\lambda _4 \mathrm{s}_{2 \alpha }+\lambda _5 \mathrm{c}_{\alpha }^2,\\
g\left( A A h^\pm h^\mp \right) & = 2 \left(\lambda _1-\lambda _3\right) \mathrm{c}_{\beta }^2-\lambda _4 \mathrm{s}_{2 \beta }+\lambda _5 \mathrm{s}_{\beta }^2,\\
g\left( \eta   A h^\pm H^\mp \right) & = \pm i \left[-2 \lambda _2 \mathrm{c}_{\beta }^2+\frac{1}{2} \left(\lambda _6-2 \lambda _7\right) \mathrm{s}_{\beta }^2\right],\\
g\left( \chi   A h^\pm H^\mp \right) & =2 \lambda _3 \mathrm{c}_{\beta }^2-\lambda _4 \mathrm{s}_{2 \beta }+ \frac{1}{2} \left(\lambda _6+2 \lambda _7\right) \mathrm{s}_{\beta }^2,\\
\begin{split}
g\left( h h H^\pm H^\mp \right) & = \mathrm{c}_{\alpha }^2 \left[2 \left(\lambda _1+\lambda _3\right) \mathrm{c}_{\beta }^2+\lambda _4 \mathrm{s}_{2 \beta }+\lambda _5 \mathrm{s}_{\beta }^2\right]-\lambda _4 \mathrm{s}_{2 \alpha } \mathrm{c}_{\beta }^2\\&~~~+\lambda _5 \mathrm{s}_{\alpha }^2 \mathrm{c}_{\beta }^2-\frac{1}{2}\left(\lambda_6 + 2\lambda_7 \right)\mathrm{s}_{2\alpha } \mathrm{s}_{2 \beta }+2 \lambda _8 \mathrm{s}_{\alpha }^2 \mathrm{s}_{\beta }^2 ,
\end{split}\\
\begin{split}
g\left( h H H^\pm H^\mp \right) & =-\frac{1}{2} \left[\lambda _1+\lambda _3+\lambda _4 \mathrm{s}_{2 \beta }-\lambda _8+ \left(\lambda _1+\lambda _3-\lambda _5+\lambda _8\right) \mathrm{c}_{2 \beta } \right]\mathrm{s}_{2\alpha } \\&~~~-\mathrm{c}_{2\alpha } \mathrm{c}_{\beta } \left[\lambda _4 \mathrm{c}_{\beta }+\left(\lambda _6+2 \lambda _7\right) \mathrm{s}_{\beta }\right],
\end{split}\\
\begin{split}
g\left( H H H^\pm H^\mp \right) & = \mathrm{s}_{\alpha }^2 \left[2 \left(\lambda _1+\lambda _3\right) \mathrm{c}_{\beta }^2+\lambda _4 \mathrm{s}_{2 \beta }+\lambda _5 \mathrm{s}_{\beta }^2\right]+\lambda _4 \mathrm{s}_{2 \alpha } \mathrm{c}_{\beta }^2\\&~~~+\lambda _5 \mathrm{c}_{\alpha }^2 \mathrm{c}_{\beta }^2+\frac{1}{2}\left(\lambda_6 + 2\lambda_7 \right)\mathrm{s}_{2\alpha } \mathrm{s}_{2 \beta }+2 \lambda _8 \mathrm{c}_{\alpha }^2 \mathrm{s}_{\beta }^2 ,
\end{split}\\
g\left( A A H^\pm H^\mp \right) & = 2 \left[\left(\lambda _1+\lambda _3\right) \mathrm{c}_{\beta }^4+2 \lambda _4 \mathrm{c}_{\beta }^3 \mathrm{s}_{\beta }+\frac{1}{4}\lambda _a \mathrm{s}_{2\beta }^2+\lambda _8 \mathrm{s}_{\beta }^4\right].
\end{align} 
\end{subequations}
The quartic couplings involving only the charged fields are:
\begin{subequations}
\begin{align}
g\left( h^\pm h^\pm h^\mp h^\mp \right) & = 4 \left(\lambda _1+\lambda _3\right),\\
g\left( H^\pm H^\pm H^\mp H^\mp \right) & = 4 \left[\left(\lambda _1+\lambda _3\right) \mathrm{c}_{\beta }^4+2 \lambda _4 \mathrm{c}_{\beta }^3 \mathrm{s}_{\beta }+\frac{1}{4}\lambda _a \mathrm{s}_{2\beta }^2+\lambda _8 \mathrm{s}_{\beta }^4\right],\\
g\left( h^\pm h^\pm H^\mp H^\mp \right) & = 4 \left[\left(\lambda _2+\lambda _3\right) \mathrm{c}_{\beta }^2-\frac{1}{2} \lambda _4 \mathrm{s}_{2 \beta }+\lambda _7 \mathrm{s}_{\beta }^2\right],\\
g\left( h^\pm h^\mp H^\pm H^\mp \right) & = 2 \left(\lambda _1-\lambda _2\right) \mathrm{c}_{\beta }^2-2 \lambda _4 \mathrm{s}_{2 \beta }+\left(\lambda _5+\lambda _6\right) \mathrm{s}_{\beta }^2.
\end{align} 
\end{subequations}

\section{Theory constraints}
\label{App:theory-constraints}
\setcounter{equation}{0}
We impose certain data-independent theory constraints on the models.
\subsection{Stability}

Necessary, but not sufficient, conditions for the stability of an $S_3$-symmetric 3HDM were provided in Ref.~\cite{Das:2014fea}. In Ref.~\cite{Emmanuel-Costa:2016vej}, based on the approach of Refs.~\cite{ElKaffas:2006gdt,Grzadkowski:2009bt}, necessary and sufficient conditions for models with $\lambda_4=0$ were discussed. It was later pointed out in Ref.~\cite{Faro:2019vcd} that parameterisation used in Ref.~\cite{Grzadkowski:2009bt}, and hence in Ref.~\cite{Emmanuel-Costa:2016vej}, is not correct\footnote{Namely, the complex product between two different unit spinors relied on six degrees of freedom. However, one of those can be expressed in terms of the other quantities, see section III-C of Ref.~\cite{Faro:2019vcd}. The following positivity condition for models with $\lambda_4=0$ (B.28)~\cite{Emmanuel-Costa:2016vej}
\begin{equation}
\lambda_{5}+\min \left(0, \lambda_{6}-2\left|\lambda_{7}\right|\right)>-2 \sqrt{\left(\lambda_{1}+\min \left(0,-\lambda_{2}, \lambda_{3}\right)\right) \lambda_{8}},
\end{equation}
yields an over-constrained $\lambda$ parameter space.}, and  \textit{one would arrive at a value of the potential which would be lower than what actually is possible to achieve within the space available}. 

In our case, due to the freedom of the $\lambda_4$ coupling, which breaks the O(2) symmetry, the stability conditions are rather involved. We parameterise the SU(2) doublets as
\begin{equation}\label{Eq.StabilityGeneralDoublets}
h_i = || h_i || \hat{h}_i, \quad i = \{1, 2, S \},
\end{equation}
where the norms of the spinors $|| h_i ||$ are parameterised in terms of the spherical coordinates
\begin{equation}
|| h_1 || = r \cos \gamma \sin \theta,  \qquad || h_2 || = r \sin \gamma \sin \theta, \qquad || h_S || = r \cos \theta,
\end{equation}
and $\hat{h}_i$ are unit spinors
\begin{equation}
\hat{h}_1 = \begin{pmatrix}
0 \\ 1
\end{pmatrix}, \qquad \hat{h}_2 = \begin{pmatrix}
\sin \alpha_2 \\ \cos \alpha_2 \, e^{i \beta_2}
\end{pmatrix},\qquad \hat{h}_S = e^{i \delta} 
\begin{pmatrix}
\sin \alpha_3 \\
\cos \alpha_3 \, e^{i \beta_3}
\end{pmatrix},
\end{equation}
where $r \geq 0,~\gamma \in[0, \pi / 2],~\theta \in[0, \pi / 2],$ and $\alpha_i \in[0,\pi/2],~\beta_i \in [0, 2\pi],~\delta \in [0, 2\pi].$ The positivity condition is satisfied, with only the quartic part being relevant, for
\begin{equation}
V_4 = \sum_i \lambda_i A_i \geq 0, \quad \forall\,\{\gamma,\theta,\alpha_i,\beta_i,\delta \},
\end{equation}
where 
\begin{subequations}
\begin{align}
A_1 &= \sin^4 \theta, \\
A_2 &= -\sin^2(2\gamma) \sin^4 \theta \cos^2 \alpha_2 \sin^2 \beta_2,\\
A_3 &= \left( \cos^4\gamma + \sin^4 \gamma \right) \sin^4 \theta - \frac{1}{2} \sin^2 (2\gamma) \sin^4 \theta \left[ \sin^2 \alpha_2 - \cos^2 \alpha_2 \cos (2 \beta_2) \right],\\
\begin{split}
A_4 &= \sin(2\theta) \sin^2 \theta \sin \gamma \bigg( \cos(2\gamma) \sin \alpha_2 \sin \alpha_3 \cos \delta \\ &\hspace{110pt} - \cos \alpha_2 \cos \alpha_3 \Big[ \sin^2 \gamma \cos(\beta_2 - \beta_3 - \delta)\\&\hspace{125pt} - \cos^2 \gamma \left\lbrace 2 \cos\left( \beta_2 - \beta_3 - \delta \right) +  \cos \left( \beta_2 + \beta_3 + \delta \right) \right\rbrace \Big]\bigg),
\end{split}\\
A_5 &= \frac{1}{4} \sin^2(2\theta),\\
\begin{split}
A_6 &= \frac{1}{4} \sin^2(2\theta) \bigg( \cos^2 \gamma \cos^2\alpha_3  + \sin^2 \gamma \Big[ \cos^2 \alpha_2 \cos^2 \alpha_3 \\& \hspace{70pt}+ \sin \alpha_3 \left\lbrace  \sin(2\alpha_2)\cos \alpha_3 \cos\left( \beta_2 - \beta_3 \right) + \sin^2 \alpha_2 \sin \alpha_3\right\rbrace \Big] \bigg),
\end{split}\\
\begin{split}
A_7 &= \frac{1}{2} \sin^2(2\theta) \bigg( \cos^2 \gamma \cos^2 \alpha_3 \cos\left( 2 \beta_3 + 2 \delta \right)\\&\hspace{30pt} + \sin^2 \gamma \Big[ \cos^2 \alpha_2 \cos^2 \alpha_3 \cos\left( 2\beta_2 - 2\beta_3 - 2\delta \right)\\&\hspace{30pt} + \sin \alpha_3 \left\lbrace  \sin^2\alpha_2 \sin \alpha_3 \cos(2\delta) + \sin(2 \alpha_2)  \cos \alpha_3 \cos\left( \beta_2 - \beta_3 - 2\delta \right)\right\rbrace \Big] \bigg),
\end{split}\\
A_8 &= \cos^4\theta.
\end{align}
\end{subequations}
First, we check if the necessary stability constraints are satisfied, see table~\ref{Table:ReproducedAnalyticalStability}. Next, with the help of the $\mathsf{Mathematica}$ function $\mathsf{NMinimize}$, using different algorithms, a further numerical minimisation of the potential is performed.

{\renewcommand{\arraystretch}{1.35}
\begin{table}[htb]
\caption{Reproduction of the necessary stability conditions of Ref.~\cite{Das:2014fea} in terms of the parameterisation given by~\eqref{Eq.StabilityGeneralDoublets}.}
\label{Table:ReproducedAnalyticalStability}
\begin{center}
\begin{tabular}{|c|c|c|}
\hline\hline
Conditions & $V_4  > 0$ & \begin{tabular}[c]{@{}l@{}}  Conditions \\ from Ref.~\cite{Das:2014fea} \end{tabular} \\ \hline\hline
\begin{tabular}[c]{@{}c@{}} $\gamma=\frac{\pi}{4},$ \\ $\theta = \alpha_2 = \frac{\pi}{2}$\end{tabular} & $\lambda_1$ & (4a) \\ \hline
$\theta=0$ & $\lambda_8$ & (4b) \\ \hline
\begin{tabular}[c]{@{}c@{}} $\gamma=\frac{\pi}{4},\, \theta= \frac{\pi}{2},$ \\ $\alpha_2=0,\, \beta_2=\{0, \frac{\pi}{2}\}$\end{tabular} & \begin{tabular}[c]{@{}c@{}} $\lambda_1+\lambda_3$ \\ $\lambda_1-\lambda_2$\end{tabular} & (4c) and (4d) \\ \hline
\begin{tabular}[c]{@{}c@{}} $\gamma = 0, \, \delta = 0,$ \\
$\tan \theta = \sqrt{\frac{\lambda_8}{\sqrt{\left( \lambda_1 + \lambda_3 \right)\lambda_8}}},$ \\
$\alpha_3 = \{0, \frac{\pi}{2}\},\,\beta_3 = \{0, \frac{\pi}{2}\}$ \end{tabular}  & \begin{tabular}[c]{@{}c@{}} $\lambda_5 + \mathrm{min}\left(0,\lambda_6 - 2\left| \lambda_7 \right|\right)$ \\ $+ 2 \sqrt{\left(\lambda_1+\lambda_3\right)\lambda_8}$\end{tabular} & (4e) and (4f) \\ \hline
\begin{tabular}[c]{@{}c@{}} $\theta= \frac{\pi}{4}, \, \gamma = \frac{\pi}{2},$\\ $\alpha_2 = \alpha_3 = \frac{\pi}{2},\,\delta=\{\pi,2\pi\}$ \end{tabular}  & \begin{tabular}[c]{@{}c@{}} $\lambda_1 + \lambda_3 \pm 2\lambda_4+ \lambda_5$ \\  $+ \lambda_6 + 2\lambda_7 + \lambda_8$\end{tabular} &  (4g)\\
\hline\hline
\end{tabular}
\end{center}
\end{table}}

\subsection{Unitarity}
\label{app:unitarity}

The tree-level unitarity conditions for the $S_3$-symemtric 3HDM were presented in Ref.~\cite{Das:2014fea}. The unitarity limit is evaluated enforcing the absolute values of the eigenvalues $\Lambda_i$ of the scattering matrix  to be within a specific limit. In our scan we assume that one is given by the value $|\Lambda_i| \leq 16\pi$~\cite{Lee:1977eg}. Some authors prefer a more severe bound $|\Lambda_i| \leq 8\pi$~\cite{Luscher:1988gc, Marciano:1989ns}. We compare the impact of both in figures~\ref{Fig:R-II-1a-masses-HA_and_charged-th_constraints}.

\subsection{Perturbativity}\label{App:Perturbativity}

The perturbativity check is split into two parts: couplings are assumed to be within the limit $|\lambda_i| \leq 4\pi$, the overall strength of the quartic scalar-scalar interactions is limited by $|g_{\varphi_i \varphi_j \varphi_k \varphi_l}| \leq4 \pi$. 

For the R-II-1a model, the list of the quartic scalar interactions $g_{\varphi_i \varphi_j \varphi_k \varphi_l}$ can be found in appendix~\ref{App:Scalar_Couplings_RII1a}. From the interactions $\eta\eta\eta\eta$ and $ \chi\chi\chi\chi$~\eqref{Eq.R_II_1a_QCLim_l1l3}, it follows that $0 < \lambda_1 + \lambda _3\leq 2 \pi/3$. Evaluation of other couplings is more involved.


\section{Supplementary equations}
\label{App:Additional}
\setcounter{equation}{0}

\subsection{Di-photon decays}\label{App:DiGamma}

The one-loop spin-dependent functions are
\begin{subequations}
\begin{align}
\mathcal{F}_{1} &=2+3 \tau+3 \tau(2-\tau) f(\tau), \\
\mathcal{F}_{1 / 2}^i &= \left\{ \begin{aligned}
& -2 \tau[1+(1-\tau) f(\tau)], & i = S,\\
& -2 \tau f(\tau), & i = P,
\end{aligned}\right. \label{Eq:F_12_S}\\
\mathcal{F}_{0} &=\tau[1-\tau f(\tau)],
\end{align}
\end{subequations}
where
\begin{equation}
\tau_i = \frac{4 m_i^2}{m_{h}^2},
\end{equation}
and 
\begin{equation}
f(\tau)=\left\{
\begin{aligned}
&\arcsin ^{2}\left(\frac{1}{\sqrt{\tau}}\right), & \tau \geq 1, \\
&-\frac{1}{4}\left[\ln \left(\frac{1+\sqrt{1-\tau}}{1-\sqrt{1-\tau}}\right)-i \pi\right]^{2}, & \tau<1.
\end{aligned}\right.
\end{equation}

\subsection{\boldmath$V$ and \boldmath$U$ matrices }\label{App:Peskin_Takeuchi_Rot}
From Refs.~\cite{Grimus:2007if,Grimus:2008nb} we determine the $V$ and $U$ matrices\footnote{Note that ``$U$'' here should not be confused with the electroweak precision observable ``$U$''.}  for R-II-1a in the Higgs basis~\eqref{RII1aHB}
\begin{subequations}
\begin{equation}
 \begin{pmatrix}
\sin(\alpha+\beta)\,h + \cos(\alpha+\beta)\,H  + i G^0 \\
 -\cos(\alpha+\beta)\,h + \sin(\alpha+\beta)\,H  + i A \\
\eta + i \chi
\end{pmatrix} = V \begin{pmatrix}
G^0 \\
A \\
h \\
H \\
\eta \\
\chi \\
\end{pmatrix},
\end{equation}
with
\begin{equation}
V = \begin{pmatrix}
i & 0 & \sin(\alpha+\beta) & \cos(\alpha+\beta) & 0 & 0 \\
0 & i &-\cos(\alpha+\beta) & \sin(\alpha+\beta) & 0 & 0 \\
0 & 0 & 0 & 0 & 1 & i
\end{pmatrix},
\end{equation}
and
\begin{equation}
 \begin{pmatrix}
G^+ \\
H^+ \\
h^+ \\
\end{pmatrix} = U \begin{pmatrix}
G^+ \\
H^+ \\
h^+ \\
\end{pmatrix}, 
\text{ with } U=\mathcal{I}_3.
\end{equation}
\end{subequations}

\clearpage

\bibliographystyle{JHEP}

\bibliography{ref}

\end{document}